\documentclass[10pt,journal,compsoc]{IEEEtran}
\ifCLASSOPTIONcompsoc
  \usepackage[nocompress]{cite}
\else
  \usepackage{cite}
\fi
\ifCLASSINFOpdf
\else
\fi
\usepackage{cite}

\usepackage{color}
\usepackage[dvipsnames]{xcolor}
\usepackage{graphicx}
\usepackage{subfigure}
\usepackage{xspace}
\usepackage{verbatim}
\usepackage{amsmath}
\usepackage{amssymb}
\usepackage{url}
\usepackage{amsthm}
\usepackage[ruled,vlined,linesnumbered]{algorithm2e}
\usepackage{tikz}
\usepackage{pgfplots}
\usepackage{tikzscale}
\usetikzlibrary{calc,positioning}
\usepackage{lipsum,adjustbox}
\pgfplotsset{compat=1.12}
\pgfplotsset{every axis legend/.append style={at={(0.5,1.03)},anchor=south},}

\begin{document}
%
\title{Edit Distance between Merge Trees}

\author{Raghavendra~Sridharamurthy,~\IEEEmembership{Student Member,~IEEE},
        Talha Bin Masood,
	    Adhitya Kamakshidasan,
        and~Vijay Natarajan,~\IEEEmembership{Member,~IEEE}
\IEEEcompsocitemizethanks{\IEEEcompsocthanksitem R. Sridharamurthy, T.B. Masood, A. Kamakshidasan, and V. Natarajan are with the Department of Computer Science and Automation, Indian Institute of Science, Bangalore, 560012.\protect\\
E-mail: \{raghavendrag,talha,adhitya,vijayn\}@iisc.ac.in}}

%
%

\markboth{{Appeared in} IEEE TRANSACTIONS ON VISUALIZATION AND COMPUTER GRAPHICS, DOI:10.1109/TVCG.2018.2873612}%
{Sridharamurthy \MakeLowercase{\textit{et al.}}: Edit Distance between Merge Trees}
%



\IEEEtitleabstractindextext{%
\begin{abstract}
Topological structures such as the merge tree provide an abstract and succinct representation of scalar fields. They facilitate effective visualization and interactive exploration of feature-rich data. A merge tree captures the topology of sub-level and super-level sets in a scalar field. Estimating the similarity between merge trees is an important problem with applications to feature-directed visualization of time-varying data. We present an approach based on tree edit distance to compare merge trees. The comparison measure satisfies metric properties, it can be computed efficiently, and the cost model for the edit operations is both intuitive and captures well-known properties of merge trees. Experimental results on time-varying scalar fields, 3D cryo electron microscopy data, shape data, and various synthetic datasets show the utility of the edit distance towards a feature-driven analysis of scalar fields.
\end{abstract}

}

\maketitle

\IEEEdisplaynontitleabstractindextext

%
\IEEEpeerreviewmaketitle

\IEEEraisesectionheading{\section{Introduction}\label{sec:introduction}}

\begin{figure}
\vspace{-0.5in}
\subfigure[2D scalar field]{\label{fig:domain}\includegraphics[height=1.3in]{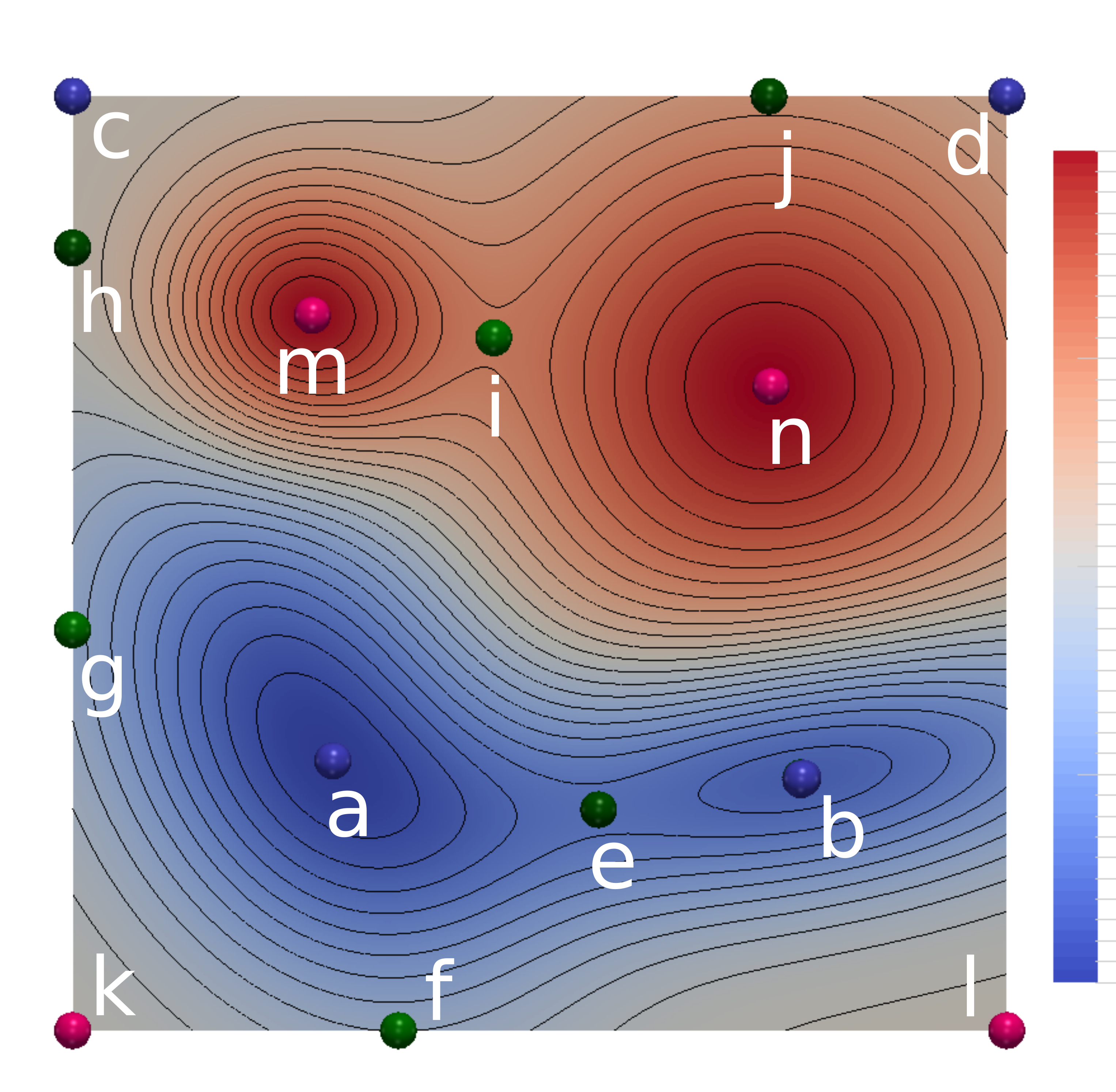}}
\subfigure[left: join tree and right: split tree]{\label{fig:join}\includegraphics[height=1.3in]{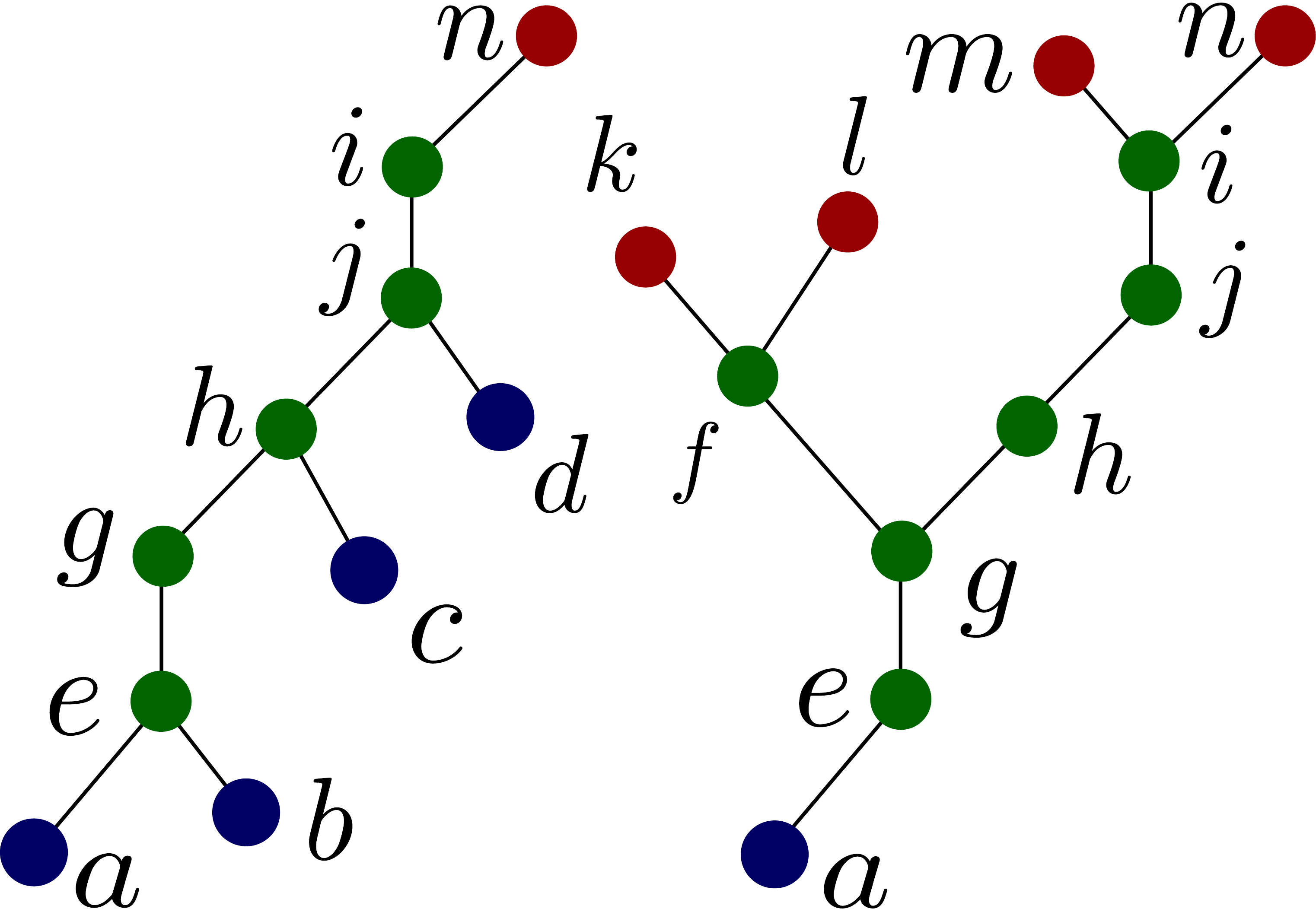}}

\caption{Merge trees. (a)~A 2D scalar field (b)~A merge tree tracks the connectivity of sub-level sets (preimage of $f^{-1} (-\infty, c]$) or the super-level sets (preimage of $f^{-1} [c,\infty)$).}
\label{fig:mergetree}
\vspace{-0.15in}
\end{figure}

\IEEEPARstart{T}{he} study of the behavior of physical quantities over time helps in understanding underlying scientific processes. Physical quantities are either measured using imaging devices or computed via simulation. In either case, they are often modeled as scalar functions (also referred to as scalar fields). 
Direct analysis and visualization of such a scalar function using isosurfaces or volume rendering provides a good overview but is limited by two factors. First, increasing size of data makes storage and retrieval inefficient. Second, the analysis often requires a sweep over a large subset of the domain or range of the function even when the features of interest may be contained within a small region. These limitations are amplified when we consider time-varying scalar functions. Thus, these techniques are not well suited for feature directed analysis and visualization. Topological structures such as the \emph{merge tree}~\cite{carr2003}  shown in Figure~\ref{fig:mergetree} provide a succinct representation of the scalar function, support feature-directed visualization and exploration, and hence enable the user to quickly identify patterns and gain insights.
Multiple scenarios demand a method for comparing scalar functions. For example, a distance or similarity measure between scalar functions is essential for detecting periodicity in a time-varying dataset. The matrix of distances between all pairs of time steps will display a characteristic pattern if the function is periodic. 
A method for comparing scalar functions is also useful for 
tracking features in time-varying phenomena~\cite{saikia2017}, topological shape matching~\cite{hilaga2001}, detecting symmetry/asymmetry in scalar fields~\cite{thomas2011,thomas2013,masood2013,Saikia2014,thomas2014}, or clustering~\cite{Oesterling2017}, computing temporal summaries of large data sets, to identify features that are preserved in ensemble simulations or multi-field data, or to compare simulated data against measured data~\cite{sauber2006multifield,nagaraj2011gradient,demir2014multi,dutta2017homogeneity}. In the above-mentioned scenarios, the similarity or dissimilarity between scalar functions is often captured by a distance measure between topological structures that represent the functions. We want such a distance measure to satisfy useful theoretical properties and be efficiently computable in order to be applicable in practice. 

\subsection{Related Work}
Assuming identical domains, RMS distance, Chebyschev distance, and other norms such as $L_p, 1 \le p \le \infty$ can be used for point-to-point comparisons. However, a direct comparison of the two scalar functions may not be appropriate because of its sensitivity to noise and minor perturbations. 

\begin{figure}
\centering
\includegraphics[width=0.40\textwidth]{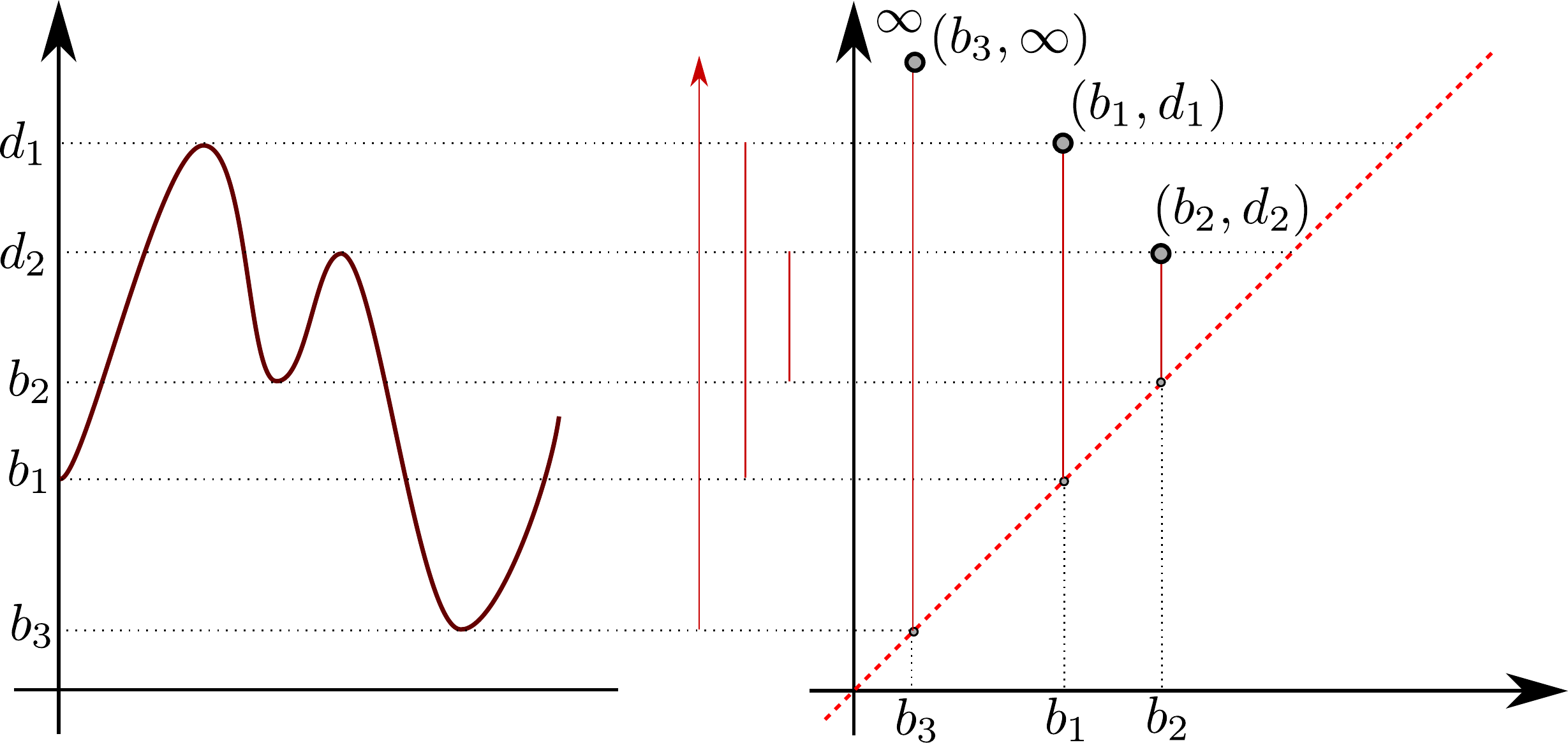}     
\caption{A 1D scalar function (left) and the persistence diagram of the function (right). Each birth-death pair $(b_i,d_i)$ is a feature of the scalar function and its persistence is defined as $d_i - b_i$. Each pair is represented as a point in $\mathbb{R}^2$. }
\label{fig:pdiag}
\vspace{-0.2in}
\end{figure}

Distance measures between various topological structures have been  studied in the literature, beginning with the bottleneck distance between persistence diagrams~\cite{cohen2007}.  A topological feature is often represented by a creator-destructor pair, a critical point pair in the case of scalar functions. For example, a minimum creates a 0-dimensional topological feature (connected component) in the sub-level set that is destroyed by a saddle. A persistence diagram (see Figure~\ref{fig:pdiag}) depicts the persistence or  ``lifetime'' of all topological features by plotting their orresponding time of creation (birth) and destruction (death) as points in $\mathbb{R}^2$. The bottleneck distance ($D_B$) between two persistence diagrams is equal to the weight of the minimum weight mapping between points of the two diagrams. The weight of a mapping is equal to the largest  $L_{\infty}$ distance between a point and its image under the mapping. We say that the persistence diagram is \emph{stable} with respect to a distance measure if it is bounded above by the $L_\infty$ distance between the two scalar functions. Intuitively, we require that small perturbations to the scalar functions translate to small changes in the distance between the respective persistence diagrams. The persistence diagram is stable with respect to $D_B$. But, the persistence diagram is only a multiset. It does not capture the spatial configuration of critical points, which reduces its discriminative capability. Morozov et al.~\cite{morozov2013} proposed the interleaving distance between merge trees. This distance is defined by a continuous map that shifts points of one merge tree onto the other and vice-versa. The distance is equal to the smallest value of the shift such that the map satisfies certain compatibility conditions. The merge tree is stable under this distance and the distance measure is more discriminative compared to the bottleneck distance but computing it is not a tractable problem. Beketayev et al.~\cite{beketayev2014} define a distance measure between merge trees that can be computed by considering all possible branch decompositions. This measure can be computed in polynomial time but provide no guarantees on stability.

Distance measures have been also defined for other topological structures. The \emph{Reeb graph} captures the topology of both sub-level sets and super-level sets of scalar functions defined on a manifold~\cite{reeb1946}. Bauer et al.~\cite{bauer2014} imposed a metric on Reeb graphs called the functional distortion distance. They proved its stability and connections with other distances such as the bottleneck and interleaving distance. The computation depends on the Gromov-Hausdorff distance, which is proven to be NP-hard~\cite{agarwal2015} to even approximate up to a constant factor for general metric graphs. Di Fabio and Landi~\cite{di2016} defined an edit distance for Reeb graphs on surfaces. They also proved its stability and showed connections with interleaving distance and function distortion distance but there is no polynomial time algorithm to compute the distance. Dey et al.~\cite{dey2015} defined the persistence distortion distance to compare metric graphs, proved its stability, and described a polynomial time algorithm with asymptotic running time $O(m^{12} \log n)$ (continuous version) and $O(n^2 m^{1.5} \log m)$ (discrete version), where $m$ is the number of edges and $n$ is the number of vertices in the larger graph. They also reported applications to shape matching.

Narayanan et al.~\cite{narayanan2015} defined a distance measure to compare \emph{extremum graphs}, whose nodes corresponds to critical points of the scalar function and arcs correspond to integral lines. The distance measure is based on the maximum weight common subgraph and they use pruning techniques to speedup the computation. While stability is not guaranteed, they present many experimental results on time-varying data to demonstrate its application to time-varying data analysis and visualization.

In contrast to the rigorous definitions of distance measures introduced in the above-mentioned works, simpler but practical similarity measures have also been studied. Saikia et al.~\cite{Saikia2014} introduced the extended branch decomposition graph (eBDG) that describes a hierarchical representation of all subtrees of a join/split tree and designed an efficient algorithm to compare them. They also present experimental results on time-varying data. Saikia et al.~\cite{saikia2015} studied a measure that compared  histograms that are constructed together with the merge trees. As in the case of bottleneck distance, this measure ignores the structure but it can be computed efficiently and is therefore useful in practice. Saikia and Weinkauf~\cite{saikia2017} later extended this measure and demonstrated applications to feature tracking in time-varying data.  

Edit distances and alignment distances for trees are inspired by edit distances defined on strings. They have found various applications,  such as comparing neuronal trees~\cite{Gillette2015}, comparing shapes~\cite{klein2000}, comparing music genre taxonomy~\cite{McVicar2016}, analysis of glycan structures~\cite{Fukagawa2011}, comparing RNA structures~\cite{Schirmer2013}, and comparing plant architectures~\cite{Ferraro2000}. Given two strings, one is transformed into the other via a sequence of operations where each operation has a non-negative associated cost. The distance is defined as the minimum cost over all such transformations. Similar distance measure may be defined for labeled trees with edit operations like relabeling, addition, and deletion of nodes. Zhang and Shasha~\cite{Zhang1989} described an algorithm to compute the tree edit distance for ordered labeled trees. Later, Zhang~\cite{Zhang1995} proposed a new algorithm for constrained tree edit distance for ordered labeled trees. The computation of tree edit distance for unordered labeled trees is NP-complete~\cite{Zhang1992}. However, the constrained version of the problem can be solved in polynomial time using a dynamic programming based algorithm~\cite{Zhang1996}. A gap corresponds to a collection of nodes that are inserted / deleted during a sequence of edit operations. Edit distance with arbitrary gap costs were first proposed by Touzet~\cite{Touzet2003}, who showed that the distance computation is NP-hard. But, the distance between ordered labeled binary trees can be computed in polynomial time~\cite{Xu2015}.

While tree edit distance based algorithms have been employed in many applications, they have not been well studied for comparing topological structures like merge trees except in very recent work. Riecke et al.~\cite{rieck2017} defined a hierarchy of persistence pairs and a tree edit distance based dissimilarity measure to compare hierarchies. Sridharamurthy et al.~\cite{Sridh2017} adapt Xu's algorithm~\cite{Xu2015}  for computing distance between ordered labeled binary trees to the case of the general subtree gap model that preserve the merge tree structure. The general subtree gap model allows for interior nodes to be inserted / deleted while retaining the child nodes. The cost model is intuitive and they present preliminary experimental results to show its utility. However, this method has multiple shortcomings:
\begin{itemize}
\item The gap model is too general. In the case of merge trees, we require a constrained version that considers gaps as persistence pairs in order to preserve the structural integrity of the tree. These pairs depend on function values. So, the constraints are ad hoc, difficult to express directly and to incorporate into the dynamic programming based algorithm that is used to compute the measure.
\item The above-mentioned pairs are not stable under perturbations to the scalar function.
\item Merge trees constructed on real world data are not necessarily binary trees.
\item Absence of a natural left-to-right ordering of children of a node in the merge tree. The algorithm requires such an ordering, random or canonical orderings lead to instabilities. 
\item The running time of the algorithm is approximately $O(n^5)$,  where $n$ is the number of nodes in the tree. This is very slow for practical applications.
\end{itemize}

In summary, existing work either propose a rigorous definition of distance with theoretical guarantees but without practical value (with very few exceptions) or describe a similarity / dissimilarity measure with practical applications but without theoretical analysis. Further, existing methods do not provide natural support for a fine grained analysis of similar/dissimilar regions.

\subsection{Contributions}
In this paper, we propose a tree edit distance based approach to compare merge trees. The distance measure is an adaptation of the constrained unordered tree edit distance~\cite{Zhang1996}, but is a significant modification that caters to merge trees and alleviates the shortcomings of the measure proposed by Sridharamurthy et al.~\cite{Sridh2017}. Individual edits correspond to topological features. The edit operations may be subsequently studied for a fine grained analysis. The paper makes the following key contributions: 
\begin{enumerate}
\item An intuitive and mathematically sound cost model for the individual edit operations.
\item A proof that the distance measure is a metric under the proposed cost model.
\item A computational solution to handle instabilities.
\item Experiments to demonstrate the practical value of the distance measure using various applications --- 2D time-varying data analysis by detecting periodicity, summarization to support visualization of 3D time-varying data, detection of symmetry and asymmetry in scalar fields, study of topological effects of subsampling and smoothing, and shape matching.
\end{enumerate}
In addition, we describe a comprehensive set of validation experiments that are designed to help understand the properties of the measure. 

\section{Background}
In this section, we introduce necessary definitions and background on merge trees, list some desirable properties of distance measures, and describe three edit operations on a merge tree that define a tree edit distance.

\subsection{Merge tree}\label{mt}
\begin{figure}

\centering
\includegraphics[width=0.2\textwidth]{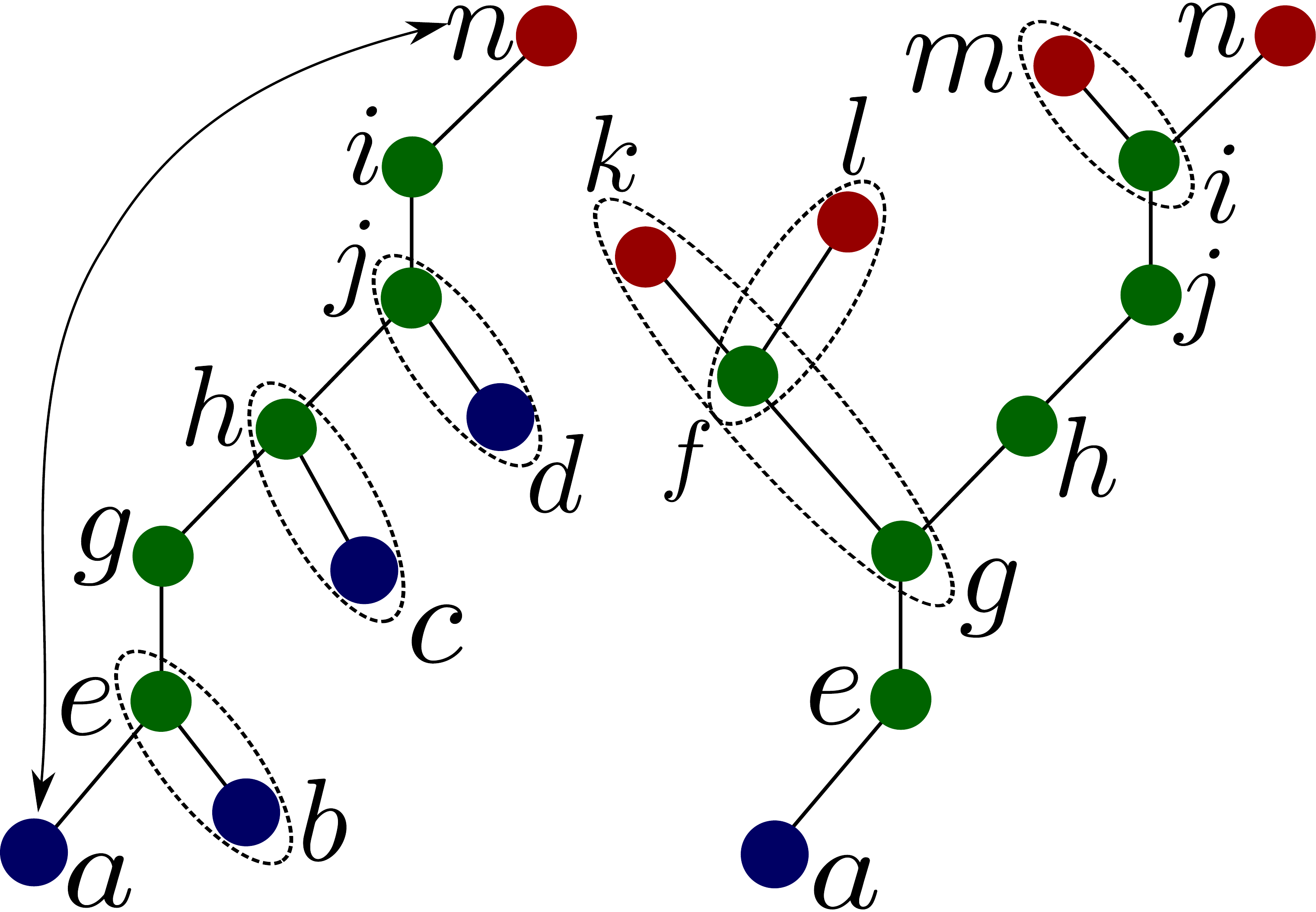}     
\caption{Persistence pairs in the join (left) and split (right) trees.}
\label{fig:jointree}
\vspace{-0.2in}
\end{figure}
A merge tree~\cite{carr2003} captures the connectivity of sub-level sets (\emph{join tree}) or super-level sets (\emph{split tree}) of a scalar function $f: \mathbb{X} \longrightarrow \mathbb{R}$ defined on a manifold domain $\mathbb{X}$, see Figure~\ref{fig:mergetree}. A value $c$ in the range of $f$ is called an \emph{isovalue}. Given an isovalue, an \emph{isocontour} is defined as the collection of all points $ x \in \mathbb{X}$ such that $f(x) = c$. 
Nodes of join trees consist of minima $M = \{m_i\}$, saddles $S = \{s_j\}$, and the global maximum. In theory, the structure of a join tree is simple. Excluding the global maximum, which is the root of the tree, every node has either 0 (minimum) or 2 children (saddle). All minima are paired with saddles based on the notion of topological persistence~\cite{edelsbrunner2000} except for one which is paired to the lone global maximum. Each such pair $(m,s)$ represents a topological feature and its persistence is defined as  $ pers(m) = pers(s) = f(s) - f(m)$. In practice, saddles may have more than two children. We discuss how to handle them in Section~\ref{hi}. A split tree is defined likewise. It contains a set of maxima and saddles together with the global minimum. Figure~\ref{fig:jointree} shows the persistence pairing for the trees from Figure~\ref{fig:mergetree}. Several fast algorithms have been developed for computing merge trees (join or split) for piecewise linear functions defined on simply connected domains~\cite{carr2003,Morozov2013dist,acharya2015parallel}.

\subsection{Distance measures}

\begin{figure}
\centering
\includegraphics[width=0.48\textwidth]{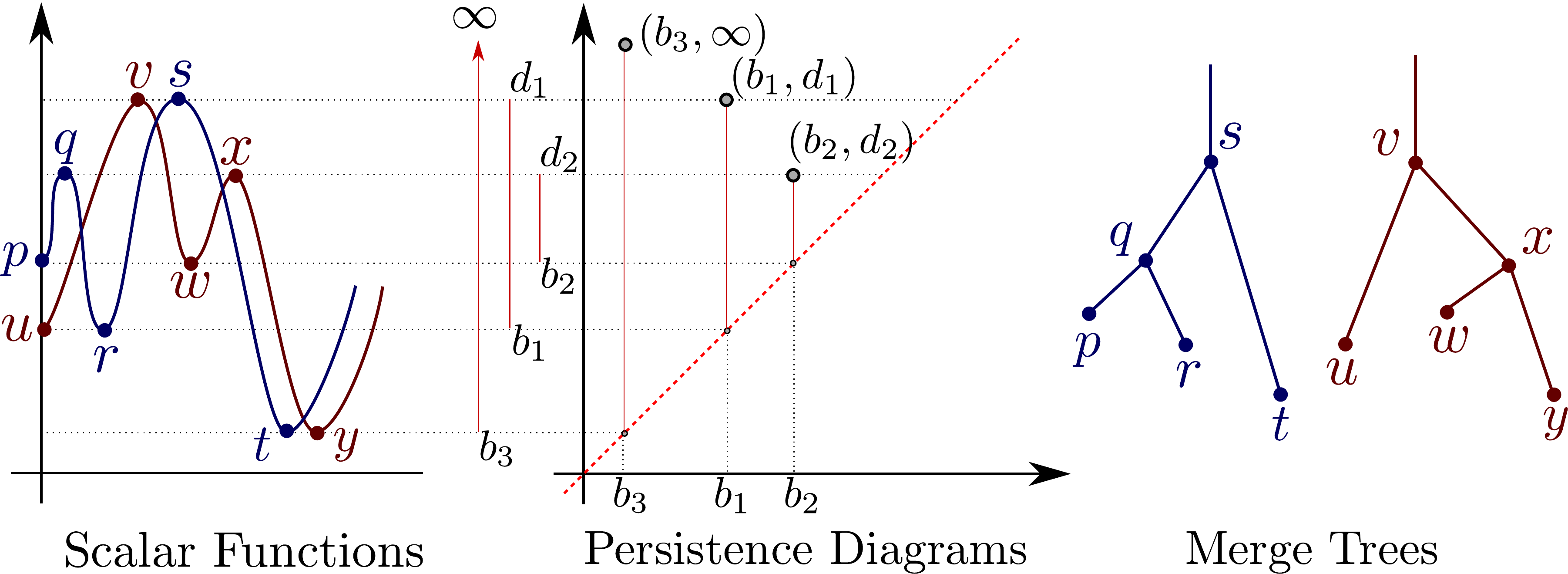}
\caption{The discriminative power of the bottleneck distance $D_B$ is low. Two scalar functions (blue and red) and the corresponding persistence diagrams and merge trees. Even though the scalar functions are different, $D_B$ is not able to capture the difference because the persistence diagrams are equal. A distance measure that considers the structure of the merge tree would discriminate the two scalar functions.}
\label{fig:discri}
\vspace{-0.2in}
\end{figure}

Designing distance measures is a well studied problem and has several applications in data analysis, visualization, pattern recognition, data mining, and machine learning.
A distance measure $D: \mathbb{X} \times \mathbb{X} \longrightarrow \mathbb{R}$ on a domain  $\mathbb{X}$ satisfies the metric properties:
\begin{enumerate}
\item \emph{Non-negativity}: $D(x,y) \geq 0$
\item \emph{Identity of indiscernibles}: $D(x,y) = 0$ iff $x = y$ 
\item \emph{Symmetry}:  $D(x,y) = D(y,x)$ 
\item \emph{Triangle inequality}: $D(x,z) \leq D(x,y) + D(y,z)$
\end{enumerate}
When metric properties such as triangle inequality are relaxed, we get a \emph{dissimilarity measure} rather than a distance measure. Further, while comparing scalar fields or topological structures constructed based on scalar fields, it is desirable that the distance measure satisfies two additional properties -- stability and discrimination. In the following discussion, we will use $D$ to refer to the distance between topological structures that represent the functions. Given two scalar functions $f,g$,
\begin{enumerate}
\item \emph{Stability}: $D(f,g) \leq \|f-g\|_\infty$
\item \emph{Discrimination}: $D_{B}(f,g) \leq D(f,g)$. 
\end{enumerate} 
Intuitively, stability requires that if the functions are not too ``different''  in terms of the $L_\infty$ norm of the difference between the functions then the distance measure between the topological structures representing the scalar functions should also be small. 
Discrimination, on the other hand, requires that however ``small" the difference between two functions, it should be captured by the distance measure. Specifically, the distance measure equals $0$ should imply that the functions are equal. Since the bottleneck distance $D_B$ between persistence diagrams of $f$ and $g$ was among the first measures defined between topological structures, we typically state this property in terms of how the distance $D$ is related to bottleneck distance.  Figure~\ref{fig:discri} shows an example where $D_B = 0$ for a pair of functions that are not equal, which implies that $D_B$ is not discriminative enough. One reason for the low discriminative power is that the persistence diagram, and hence $D_B$, does not incorporate the connectivity between critical points as in the merge tree.  We wish to design a distance measure that satisfies the following property:
\begin{align}
D_{B}(f,g) \leq D(f,g) \leq \|f-g\|_\infty
\end{align}
From the computational perspective, $D$ should be computable either exactly or within a constant factor of approximation in polynomial time in order for it to be useful in a practical application.

\subsection{Tree edit distance}
Tree edit distances have been studied extensively in the past few decades~\cite{Bille2005}. All these measures typically employ a set of edit operations with associated costs and try to minimize the total cost over the set of all edit operations. Let $T$ be a rooted tree with node set $V$ and edge set $E$. For a node $v \in V$, $deg(v)$ is the number of children of $v$,  and $parent(v)$ is its parent in the tree. The maximum degree of a node in the tree is denoted as $deg(T)$. We denote an empty tree by $\theta$. Since we are interested in labeled trees, let $\Sigma$ be the set of labels, and $\lambda \notin \Sigma$ denote the null or empty character, which corresponds to a gap. In the following discussion, we use notations and definitions from Zhang~\cite{Zhang1996}.

\begin{figure}
\centering

\subfigure[delete]{\label{fig:delete}\includegraphics[width=0.23\textwidth]{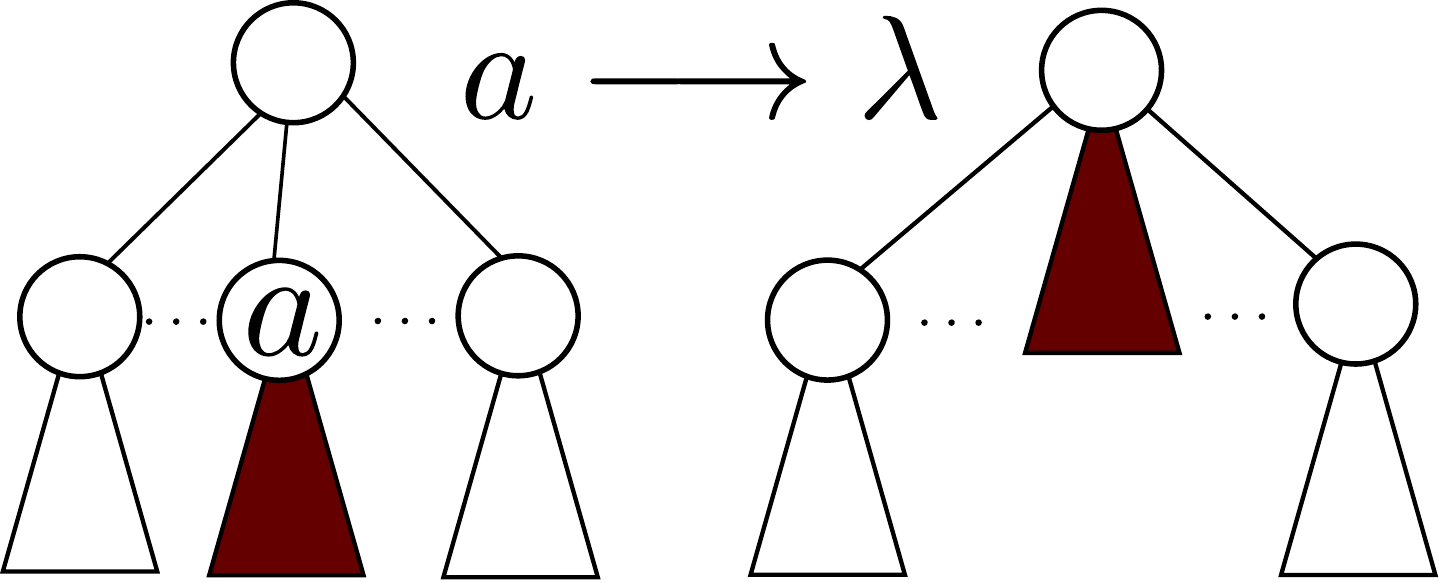}}
~
\subfigure[insert]{\label{fig:insert}\includegraphics[width=0.23\textwidth]{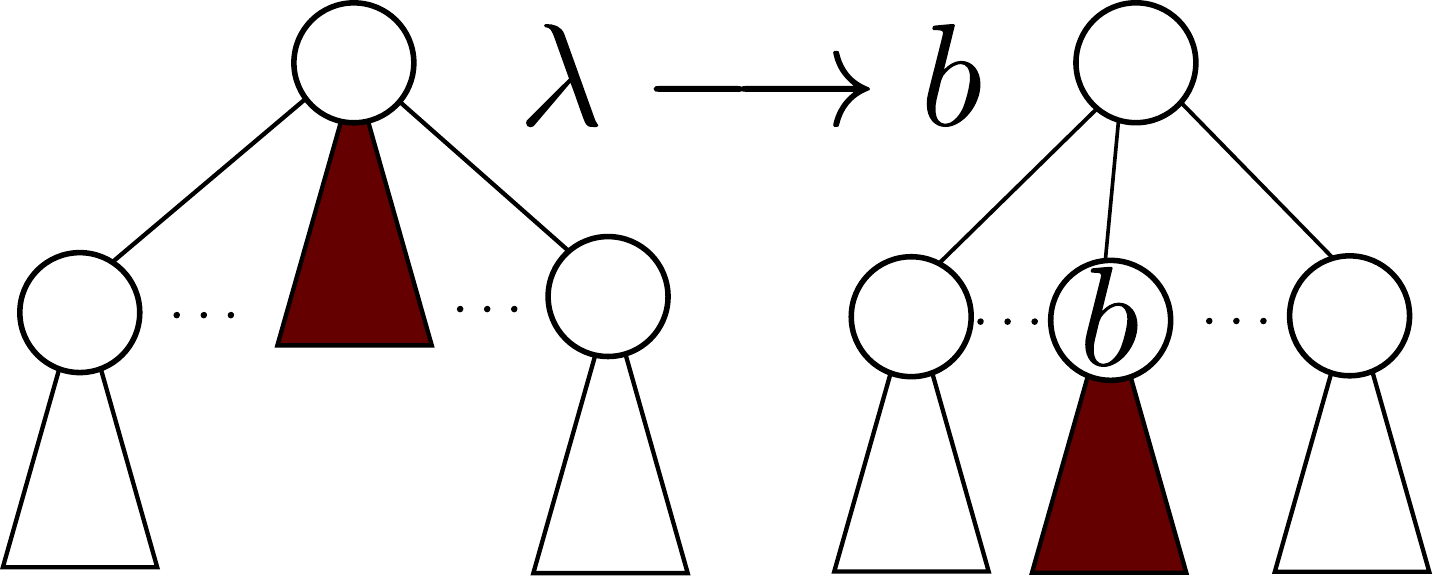}}

\subfigure[relabel]{\label{fig:relabel}\includegraphics[width=0.23\textwidth]{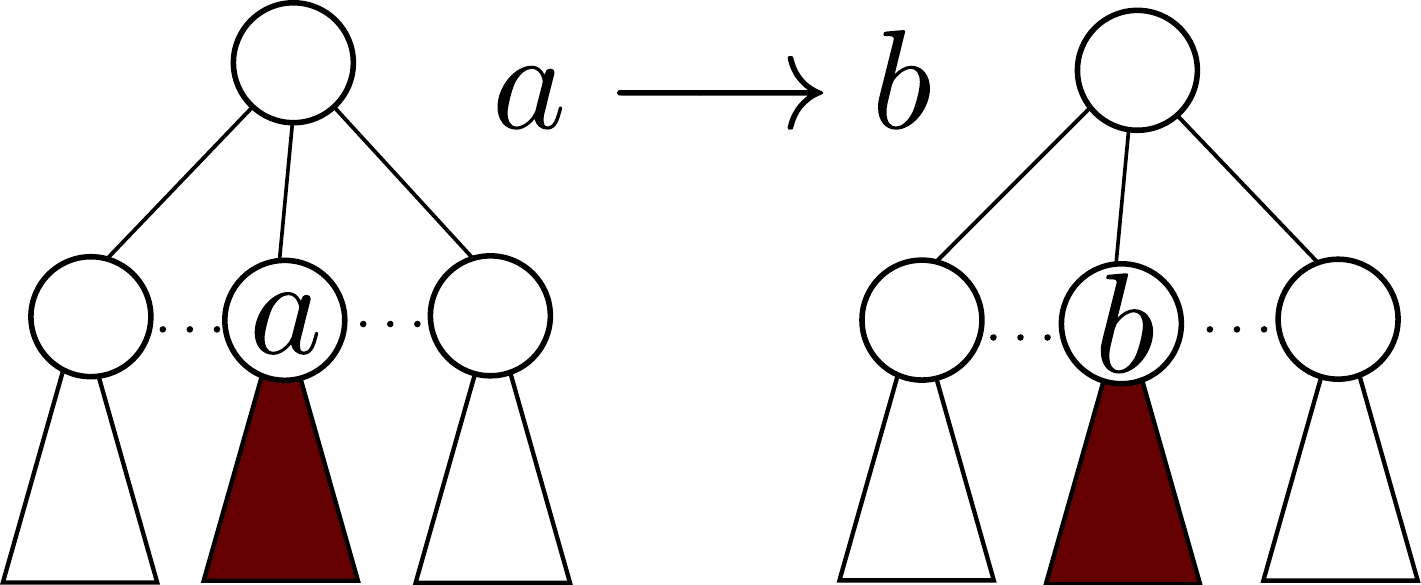}}
\caption{Three different tree edit operations. Each edit affects only one node in the tree. The null character $\lambda$ corresponds to a gap.}
\label{fig:oper}
\vspace{-0.2in}
\end{figure}

\noindent\textbf{Edit operations.}\label{cost}  
The edit operations differ based on the gap model. For this discussion we consider edit operations that modify the tree, one node at a time. Xu~\cite{Xu2015} gives a detailed discussion of general gaps where edits modify multiple nodes. We consider a total of three edit operations as shown in Figure~\ref{fig:oper}.

\begin{enumerate}
\item \textbf{relabel:} A relabel $a \longrightarrow b$ corresponds to an operation where the label $a \in \Sigma$ of a node is changed to a label $b \in \Sigma$.
\item \textbf{delete:} A delete operation $a \longrightarrow \lambda$ removes a node $n$ with label $a \in \Sigma$ and all the children of $n$ are made the children of $parent(n)$. 
\item \textbf{insert:} An insert operation $\lambda \longrightarrow b$ inserts a node $n$ with label $b \in \Sigma$ as a child of another node $m$ by moving all the children of $m$ to children of $n$.
\end{enumerate}

We define a cost function $\gamma$ that assigns a non-negative real number to each edit operation of the form $a \longrightarrow b$. It is useful if the cost function $\gamma$ satisfies metric properties i.e. $\forall a,b,c \in \Sigma \cup \{\lambda\}$
\begin{enumerate}
\item $\gamma(a \longrightarrow b) \ge 0,\  \gamma(a \longrightarrow a) = 0$
\item $\gamma(a \longrightarrow b) = \gamma(b \longrightarrow a)$
\item $\gamma(a \longrightarrow c) \le \gamma(a \longrightarrow b) + \gamma(b \longrightarrow c)$
\end{enumerate}
In particular, Zhang~\cite{Zhang1996} proved that if $\gamma$ is a metric then the edit distance is also a metric, else it will be merely a dissimilarity measure. Given a tree $T_1$, we can apply a sequence of edit operations to transform it into another tree $T_2$. If $S = s_1,s_2, \ldots , s_k$ is a sequence of edit operations, where each $s_i$ is an edit, we can extend the cost function to $S$ by defining $\gamma(S) = \Sigma_{i=1}^{|S|} \gamma(s_i)$.

\noindent\textbf{Edit distance.} 
Formally, the distance between two trees $T_1, T_2$ is defined as
\begin{align}
D_e(T_1,T_2) = \min_{S}\{\gamma(S)\}
\end{align}
where $S$ is an edit operation sequence from $T_1$ to $T_2$. 

\section{Edit Distance Mappings}
Computing the edit distance between merge trees is a minimization problem with a huge search space. In order to understand this search space and how it affects the computation, we first define some edit distance mappings -- unconstrained, constrained, and restricted -- and their properties as described by Zhang~\cite{Zhang1996}. We refer the reader to the supplementary material for additional description and illustrations of the mappings.

\subsection{Unconstrained edit distance mapping}\label{uncon}
The sequence of edit operations performed to transform $T_1$ into $T_2$ determines a mapping between the two trees. For convenience, we order the nodes of both the trees. This ordering does not affect the distance. 
Let $t_1$ and $t_2$ denote the ordering of nodes in $T_1$ and $T_2$, respectively, and $t_1[i]$ represents the $i$th node in the ordering. 
Let  $M_e$ denote a collection of ordered integer pairs $(i,j)$. A triple $(M_e,T_1,T_2)$ defines the \emph{edit distance mapping} from $T_1$ to $T_2$,  where each pair $(i_1,j_1),(i_2,j_2) \in M_e$ satisfies the following properties:
\begin{itemize}
\item $i_1 = i_2$ iff $j_1 = j_2$ (one-to-one)
\item $t_1[i_1]$ is an ancestor of $t_1[i_2]$ iff $t_2[j_1]$ is an ancestor of $t_2[j_2]$ (ancestor ordering).
\end{itemize}
The cost of transforming $T_1$ into $T_2$ can be expressed through the mapping as 
\begin{eqnarray}
\gamma(M_e) & = \sum\limits_{(i,j) \in M_e} \gamma(t_1[i] \longrightarrow t_2[j]) \nonumber \\ 
&+ \sum\limits_{\{i| \nexists j, (i,j) \in M_e\}} \gamma(t_1[i] \longrightarrow \lambda) \nonumber \\ 
			& + \sum\limits_{\{j| \nexists i, (i,j) \in M_e\}} \gamma(\lambda \longrightarrow t_2[j])
\end{eqnarray}
Given a sequence of edit operations $S$ that transforms $T_1$ into $T_2$, there exists a mapping $M_e$ such that $\gamma(M_e) \le \gamma(S)$.  Conversely, given an edit distance mapping $M_e$ ,there exists a sequence of edit operations $S$ such that $\gamma(S) = \gamma(M_e)$. Using the above, it can be shown that 
\begin{align}
D_e(T_1,T_2) = \min_{M_e}\{\gamma(M_e)\}
\end{align}
where $(M_e,T_1,T_2)$ defines the \emph{edit distance mapping} from $T_1$ to $T_2$. Zhang et al.~\cite{Zhang1992} showed that computing $D_e(T_1,T_2)$ is NP-complete even when the trees are binary and $|\Sigma| = 2$. 

\subsection{Constrained and restricted mappings}\label{con}
Adding constraints to the edit distance mapping brings it within the computationally tractable realm. The main constraint imposed is that disjoint subtrees are mapped to disjoint subtrees. Let $T[i]$ denote the subtree rooted at the node with label $i$ and $F[i]$ denote the unordered forest obtained by deleting the node $t[i]$ from $T[i]$. A node $t_1[i]$ is a \emph{proper ancestor} of $t_1[j]$  if $t_1[i]$ lies on the path from the root to $t_1[j]$ and $t_1[i] \ne t_1[j]$. The triple $(M_c,T_1,T_2)$ is called a \emph{constrained edit distance mapping} if,
\begin{itemize}
\item $(M_c,T_1,T_2)$ is an edit distance mapping, and
\item Given three pairs $(i_1, j_1),(i_2, j_2),(i_3, j_3)\in M_c$, the \emph{least common ancestor} $lca(t_1[i_1], t_1[i_2])$ is a proper ancestor of $t_1[i_3]$  iff $lca(t_2[j_1], t_2[j_2])$ is a proper ancestor of $t_2[j_3]$.
\end{itemize}

The constrained edit distance mappings can be composed. Given two constrained edit distance mappings $M_{c_1}$ from $T_1$ to $T_2$ and $M_{c_2}$ from $T_2$ to $T_3$, $M_{c_2} \circ M_{c_1}$ is a constrained edit distance mapping between $T_1$ and $T_3$. Also,
\begin{align}
\gamma(M_{c_2} \circ M_{c_1}) \le \gamma(M_{c_1}) + \gamma(M_{c_2})
\end{align}
which can be proven using the triangle inequality imposed on the edit operation costs. This leads to the definition of constrained edit distance
\begin{align}
D_c(T_1,T_2) = \min_{M_c}\{\gamma(M_c)\}
\end{align}
$D_c$ also satisfies metric properties.
Both $M_e$ and $M_c$ deal with mapping between unordered trees. Similar mappings work for forests. We define a \emph{restricted mapping} $M_r(i,j)$ between $F_1[i]$ and $F_2[j]$ as follows:
\begin{itemize}
\item $M_r(i,j)$ corresponds to a constrained edit distance mapping between $F_1[i]$ and $F_2[j]$.

\item Given two pairs $ (i_1, j_1), (i_2, j_2) \in M_c $, $t_1[l_1]$ and $t_1[l_2]$ belong to a common tree in $F_1[i]$ if and only if $t_2[j_1]$ and $t_2[j_2]$ belong to a common tree in $F_2[i]$.

\end{itemize}
Essentially, nodes within different trees of $F_1$ are mapped to nodes lying in different trees of $F_2$.

\subsection{Constrained edit distance}\label{sec:prop}
We recall the properties of $D_c$.
Let $t_1[i_1], t_1[i_2], \ldots, t_1[i_{n_i}]$ be the children of $t_1[i]$ and $t_2[j_1], t_2[j_2], \ldots, t_2[j_{n_j}]$ be the children of $t_2[j]$. Further, let $\theta$ denote the empty tree. Then, 
\begin{align}
D_c(\theta,\theta) &= 0, \\
D_c(F_1[i],\theta) &= \sum\limits_{k=1}^{n_i} D_c(T_1[i_k],\theta), \\
D_c(T_1[i],\theta) &= D_c(F_1[i],\theta) + \gamma(t_1[i] \longrightarrow \lambda),\\
D_c(\theta,F_2[j]) &= \sum\limits_{k=1}^{n_j} D_c(\theta,T_2[j_k]), \\
D_c(\theta,T_2[j]) &= D_c(\theta,F_2[j]) + \gamma(\lambda \longrightarrow t_2[j]),
\end{align}
\begin{align}
&\scriptsize D_c(T_1[i],T_2[j]) \nonumber \\
&= \scriptsize \min \begin{cases}
D_c(\theta,T_2[j]) + \min\limits_{1 \le t \le n_j} \{D_c(T_1[i], T_2[j_t])- D_c(\theta, T_2[j_t])\},\\
D_c(T_1[i],\theta) + \min\limits_{1 \le s \le n_i} \{D_c(T_1[i_s], T_2[j])- D_c(T_1[i_s],\theta)\},\\
D_c(F_1[i], F_2[j]) + \gamma(t_1[i] \longrightarrow t_2[j]).
\end{cases}
\end{align}
If the cost is not a metric, we need to include one additional case, namely $D_c(F_1[i], F_2[j]) + \gamma(t_1[i] \longrightarrow \lambda) + \gamma(\lambda \longrightarrow t_2[j])$. The distance between two forests is given by
\begin{align}
&\scriptsize D_c(F_1[i],F_2[j]) \nonumber \\
&= \scriptsize \min \begin{cases}
D_c(\theta,F_2[j]) + \min\limits_{1 \le t \le n_j} \{D_c(F_1[i], F_2[j_t])- D_c(\theta, F_2[j_t])\},\\
D_c(F_1[i],\theta) + \min\limits_{1 \le s \le n_i} \{D_c(F_1[i_s], F_2[j]) - D_c(F_1[i_s],\theta)\},\\
\min\limits_{M_r(i,j)} \gamma(M_r(i,j)).
\end{cases}
\end{align}
The minimum restricted mapping may be computed by constructing a weighted bipartite graph in such a way that the cost of the minimum weight maximum matching $MM(i,j)$ is exactly the same as the cost of the minimum restricted mapping $M_r(i,j)$,
\begin{align}
\min\limits_{M_r(i,j)} \gamma(M_r(i,j)) = \min\limits_{MM(i,j)} \gamma(MM(i,j))
\end{align}

\subsection{Algorithm}\label{tedalgo}
Zhang described an algorithm for computing the tree edit distance for labeled unordered trees~\cite{Zhang1996}. It is a dynamic programming based algorithm that follows from the properties discussed in Section~\ref{sec:prop}. The pseudo code is presented in the supplementary material (Section~2). The entry $D(T_1[m], T_2[n])$ in the table with $m = |T_1|$ and $n = |T_2|$ corresponds to the final result. The  algorithm computes the distance in $O(|T_1| \times |T_2| \times (deg(T_1)+deg(T_2)) \times log_{2}(deg(T_1)+deg(T_2)))$ time in the worst case.

\section{Tree Edit Distance}
We now describe a new tree edit distance that is appropriate for comparing merge trees, discuss its properties, and an algorithm for computing the distance measure. 

\subsection{Comparing merge trees}\label{cmpmergetree}
Our proposed measure is based on a variant of tree edit distance that applies to unordered general trees as opposed to ordered binary trees. This variant is appropriate because
\begin{itemize}
\item Merge trees are unordered trees.
\item Merge trees are not binary in general.
\item Persistence pairs represent topological features. So, it is natural that the edit operations are defined in terms of persistent pairs.
\item The pairs do not fit into any subtree gap model that has been studied in the literature.
\end{itemize}

Consider the properties of edit distance mapping mentioned in Section~\ref{uncon}, but now in the context of merge trees.  The one-to-one property is applicable but ancestor ordering might not hold in all cases. Small perturbations in the function value may result in swaps similar to rotations in AVL or red-black trees~\cite[Chapter~13]{cormen2009}, which  violate the ancestor ordering. Such violations also result in instabilities \textit{i.e.}, cause significant fluctuations in the distance (see Section \ref{hi}). Computing the edit distance with the ancestor order preserving mappings is already infeasible. Removing that constraint will make the computation more difficult.  We introduce a stability parameter to ensure that  ancestor order preserving mappings are identified in practically all cases. More details on this computational solution to handling instabilities can be found in Section~\ref{hi}. This solution does discard some mappings and may lead us away from the optimum solution. But, the stabilization ensures that the mapping remains meaningful and helps reduce the search space thereby making the problem tractable. 

To summarize, $D_c$ between unordered trees with suitable modifications seems to be a good candidate for comparing merge trees. In this section, we describe one such distance measure and demonstrate its use in the following section. The additional constraint of mapping disjoint subtrees to disjoint subtrees may seem limiting. Also, $D_e(T_1,T_2) \le D_c(T_1,T_2)$, which implies that the constrained edit distance may not be optimal in many cases. But, we observe that, in practice, it is not as limiting and gives good results in many applications. 

\subsection{Cost model}\label{cm}
The edit distance mapping $M_e$ and the constrained edit distance mapping $M_c$ need to be suitably modified so that they are applicable for comparing merge trees. We begin by considering the edit operations as applicable to merge trees together with appropriate cost models. The literature on tree edit distances study generic trees and hence do not describe particular cost models. The following discussions focus on join trees but all results hold for split trees also.

\begin{figure}
\centering
\includegraphics[width=0.47\textwidth]{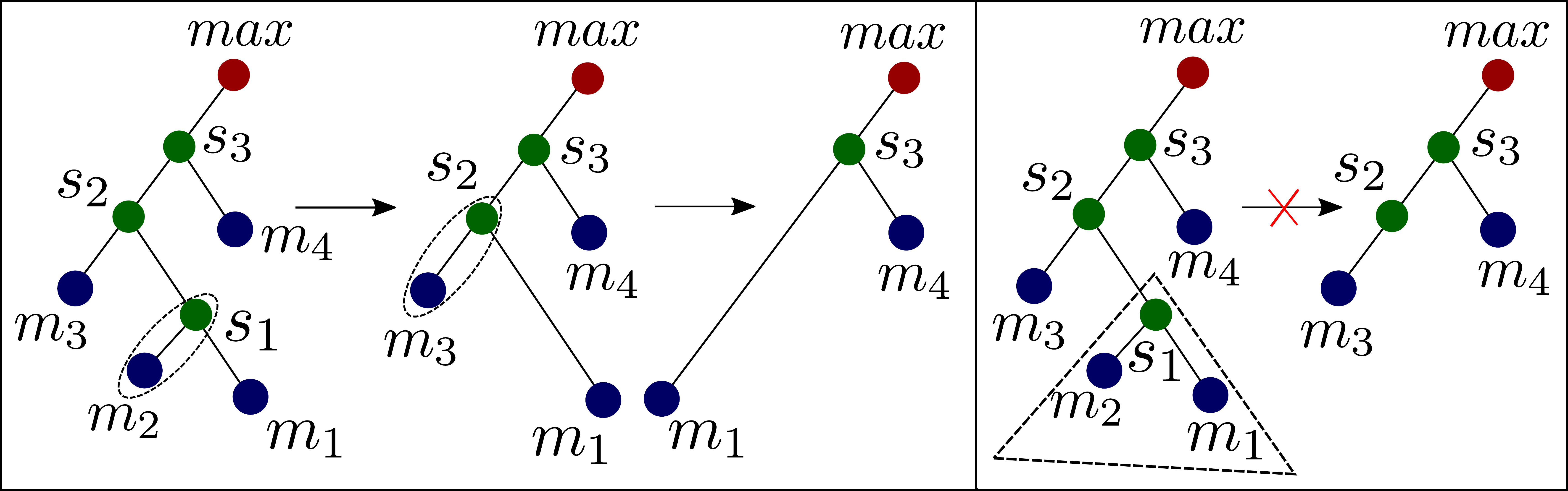}
\caption{Permitted and forbidden edit operations. (left)~A gap is introduced by removing a persistence pair. (right)~An edit operation that is permitted for generic trees but is invalid for a join tree. Nodes and arcs are repositioned to improve the tree layout.}
\label{fig:validop}
\vspace{-0.1in}
\end{figure}

Tree edit operations on the join tree need to preserve the structural integrity of the join tree. This reduces the number of operations, say insertions and deletions. Consider a min-saddle pair $(m_2,s_1)$ in Figure~\ref{fig:validop}. If $s_1$ is deleted, then it's pair $m_2$ should also be deleted, and vice-versa. After deletion, $m_1$ is adjacent to $s_2$. But deletion of $s_1$ does not necessarily require that the entire subtree rooted at $s_1$ be deleted. In fact, deleting the entire subtree may not result in a valid join tree as illustrated in Figure~\ref{fig:validop}. In this particular illustration, we consider the pairing imposed by the persistence. But, in general, we may consider other pairings based on say volume, hyper-volume, etc.

Gaps in the join tree can be represented as a collection of min-saddle pairs. In Figure~\ref{fig:validop}, we can transform the first tree into the last tree by deleting the pairs $\{(m_2,s_1),(m_3,s_2)\}$. 
We propose two cost models that capture the preservation of topological features and are applicable for join trees. Consider nodes $p \in T_1$ and $q \in T_2$. Then $p$ and $q$  are creators or destroyers of  topological features in $T_1$ and $T_2$, respectively. Let the birth and death times of these features be $(b_p, d_p)$ and $(b_q, d_q)$, respectively. These birth-death pairs correspond to points in the persistence diagrams.  Alternatively, they are represented as closed intervals $[b_p,d_p]$ and $[b_q,d_q]$ in a persistence barcode.

\subsubsection{$L_{\infty}$ cost $C_{W}$}\label{cw}
\begin{align}
\gamma(p \longrightarrow q) &= \min \begin{cases}
\max(|b_q-b_p|,|d_q-d_p|),\\
\frac{(|d_p-b_p| + |d_q-b_q|)}{2}
\end{cases}\\
\gamma(p \longrightarrow \lambda) &= \frac{|d_p-b_p|}{2}\\
\gamma(\lambda \longrightarrow q) &= \frac{|d_q-b_q|}{2}
\end{align}
This cost model is based on the bottleneck and Wasserstein distances. Note that the insert / delete cost is based on the $L_{\infty}$-distance of the points $p$ (or $q$) from the diagonal in the persistence diagram. The relabel cost is the minimum of the $L_{\infty}$-distance between the points $p$ and $q$ and the sum of the $L_{\infty}$-distance from the points $p$ (or $q$) to the diagonal. This corresponds to the scenario where transforming $p$ to $q$ ($p \longrightarrow q$) by deleting $p$ and inserting $q$ ($p \longrightarrow \lambda$ and $\lambda \longrightarrow q$) has a lower cost in some cases. Figure~\ref{fig:cost} shows how these costs can be derived from the persistence diagram when there is no overlap in the barcodes and when there is overlap between the barcodes.

\subsubsection{Overhang cost $C_O$}
\begin{figure}
\centering
\includegraphics[width=0.48\textwidth]{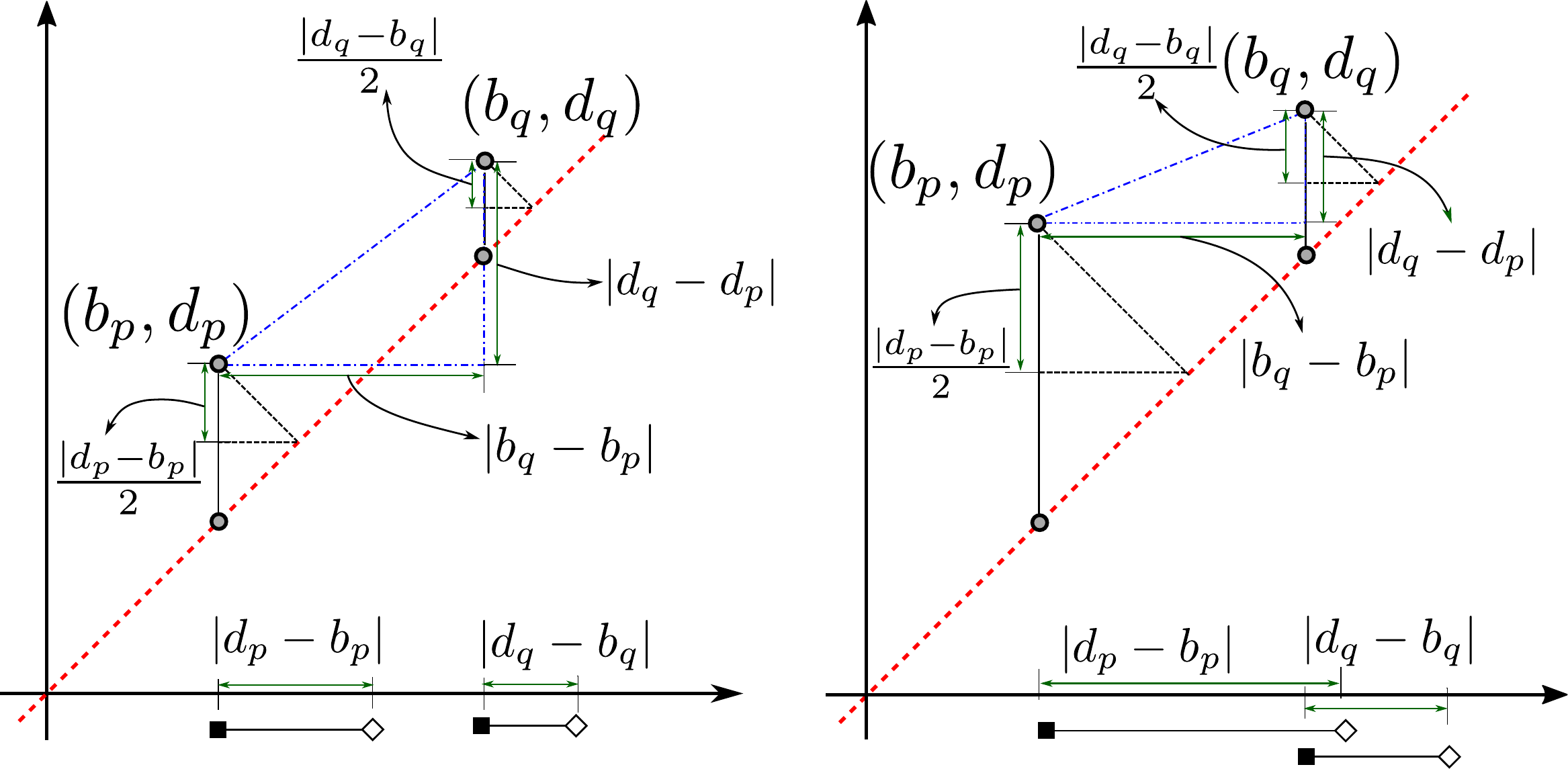}
\caption{Illustration of the cost models when there is no overlap in the barcodes (left) and when there is overlap (right). We distinguish between birth and death events in the barcode by using different glyphs for the start and the end of the intervals.}
\label{fig:cost}
\vspace{-0.1in}
\end{figure}

\begin{align}
\gamma(p \longrightarrow q) &= \min \begin{cases}
|b_q-b_p| + |d_q-d_p|,\\
|d_p-b_p| + |d_q-b_q|
\end{cases}\\
\gamma(s \longrightarrow \lambda) &= |d_p-b_p|\\
\gamma(\lambda \longrightarrow t) &= |d_q-b_q|
\end{align}
This cost model is based on the overlap of the barcodes or the intervals. We consider the lengths of the overhang or the non-overlapping section to determine the costs. Consider $p \longrightarrow \lambda$, the interval corresponding to $p$ is given by $[b_p,d_p]$ with length $|d_p-b_p|$ and the interval corresponding to $\lambda$ is $\emptyset$ with length $0$. Since there is no overlap, the cost is $|d_p-b_p| + 0 = |d_p-b_p|$. The cost of $\lambda \longrightarrow q$ can be derived similarly. Let us now consider the cost of $p \longrightarrow q$. If there is an overlap, we discard the overlap and obtain $|b_q-b_p|+|d_q-d_p|$. If there is no overlap then the cost is equal to $|d_p-b_p| + |d_q-b_q|$. The minimum of the two expressions is the relabel cost. The barcodes are shown in the fourth quadrant of the persistence diagrams in Figure~\ref{fig:cost}.

\subsection{Metric properties}
Metric property enables us to study the space of all the trees, compute the mean, and also compose transformations between merge trees. From  Sections~\ref{cost} and~\ref{con}, we know that if the cost model satisfies the metric property then the distance measure is also a metric. We now prove the metric properties for our cost model.  

The overhang cost is similar to symmetric difference, which is a  well-known metric~\cite{levandowsky1971}. We now prove that the $L_{\infty}$ cost $C_{W}$ is a metric. 

\subsubsection{$C_{W}$ is a metric}
We show that the cost $C_{W}$ is equal to the Wasserstein distance between two corresponding persistence diagrams. 
Let $N$ denote the set of all nodes in the merge trees and $\lambda$ denote a node corresponding to the null character. We define a mapping $\mathcal{M} : N \cup \{\lambda\} \longrightarrow Dgm$, where $Dgm$ is set of all persistence diagrams, as follows:
\begin{enumerate}
\item $\forall p \in N,\ \mathcal{M}(p) = \{(b_p,d_p)\} \cup \{(x,x), x \ge 0\}$, 
\item $\mathcal{M}(\lambda) = \{(x,x), x \ge 0\}$.
\end{enumerate}
Define the distance on the set $N \cup \{\lambda\}$ as the Wasserstein distance of the first order \emph{i.e.}, given $p,q \in N \cup \{\lambda\}$  
\[ d(p,q) = W_1(\mathcal{M}(p),\mathcal{M}(q)) \]
Now, the cost $C_{W}$ can be rewritten as
\begin{align}
\gamma(p \longrightarrow q) &= W_1(\mathcal{M}(p),\mathcal{M}(q))\\
\gamma(p \longrightarrow \lambda) &= W_1(\mathcal{M}(p),\mathcal{M}(\lambda))\\
\gamma(\lambda \longrightarrow q) &= W_1(\mathcal{M}(\lambda),\mathcal{M}(q)).
\end{align}
Since the Wasserstein distance $W_1(\cdot,\cdot)$ between persistence diagrams is known to be a metric~\cite[Chapter~6]{Villani2009}, $C_{W}$ is also a metric. 
However, this proof of the metric property is for general distributions. We have an alternative proof for merge trees from first principles with the aim to better understand the cost. 

\subsubsection{$C_{W}$ is a metric : proof from first principles}
Non-negativity and symmetry follows by definition because $C_{W}$ is based on sum, max, min of absolute values. \begin{figure}
\centering
\includegraphics[width=0.35\textwidth]{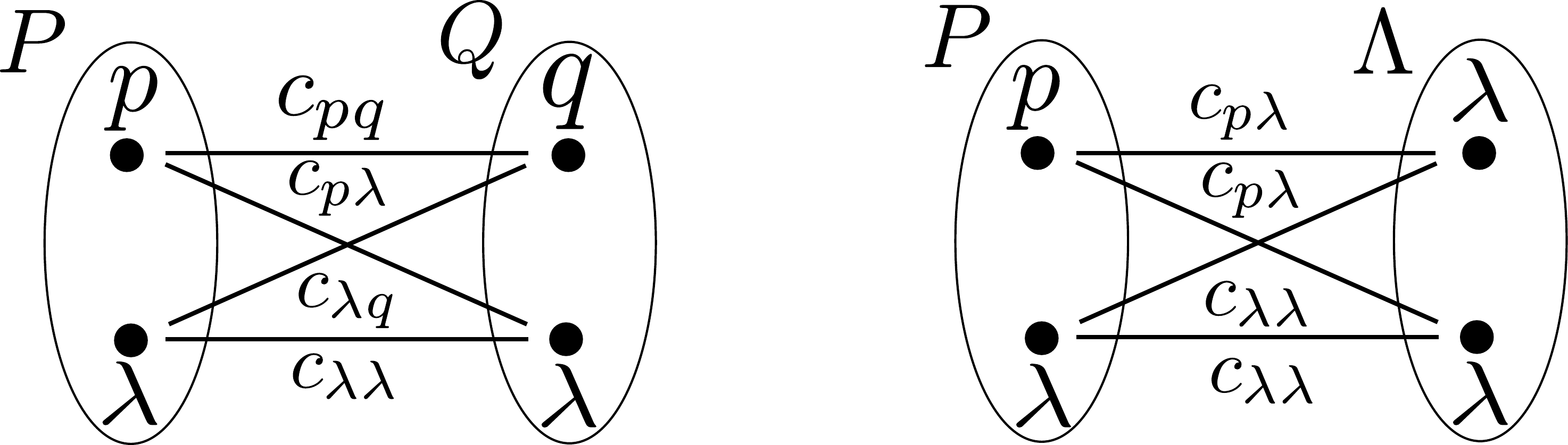}
\caption{The cost of edit operations can be reformulated as the weight of a minimum weight maximum matching in a bipartite graph. Bipartite graph for a relabel operation $p \longrightarrow q$ (left) and a delete operation~(right).}
\label{fig:bipcon}
\end{figure}

To prove the triangle inequality, we first reformulate the cost of the edit operations as the weight of a minimum weight maximum matching. The matching is defined in a bipartite graph. Nodes of the bipartite graph consists of  the merge tree nodes together with an equal number of copies of $\lambda$. We collect the nodes of the graph to construct sets of the form $P = \{p, \lambda\}$ and a special multiset $\Lambda = \{\lambda,\lambda\}$, see Figure~\ref{fig:bipcon}. All pairs of nodes from different multisets are connected by an edge. The edge weight $c$ is given by the $L_\infty$ distance between the corresponding points in the persistence diagram:
\begin{align}
c_{pq} = L_\infty(p, q) &= \max(|b_q-b_p|,|d_q-d_p|)\\
c_{p\lambda} = L_\infty(p, \lambda) &= \frac{|d_p-b_p|}{2}\\
c_{\lambda q} = L_\infty(\lambda, q) &= \frac{|d_q-b_q|}{2}\\
c_{\lambda \lambda} = L_\infty(\lambda, \lambda) &= 0
\end{align}
The cost of the edit operations is equal to the cost of the minimum weight maximum matching $MM$ in this bipartite graph. In Figure~\ref{fig:bipcon}, one of the two matchings will determine the cost of the edit operation. 
\begin{align}
\gamma(p \longrightarrow q) = MM(P,Q) &= \min \begin{cases}
c_{pq} + c_{\lambda \lambda},\\
c_{p\lambda} + c_{\lambda q}
\end{cases}\\
\gamma(p \longrightarrow \lambda) = MM(P,\Lambda) &= \min \begin{cases}c_{p\lambda} + c_{\lambda \lambda}, \\
c_{\lambda \lambda} + c_{p\lambda} 
\end{cases}\\
\gamma(\lambda \longrightarrow q) = MM(\Lambda,Q) &= \min \begin{cases}
c_{\lambda \lambda} + c_{\lambda q},\\
c_{\lambda q} + c_{\lambda \lambda} 
\end{cases}
\end{align}

\begin{figure}
\centering
\includegraphics[width=0.43\textwidth]{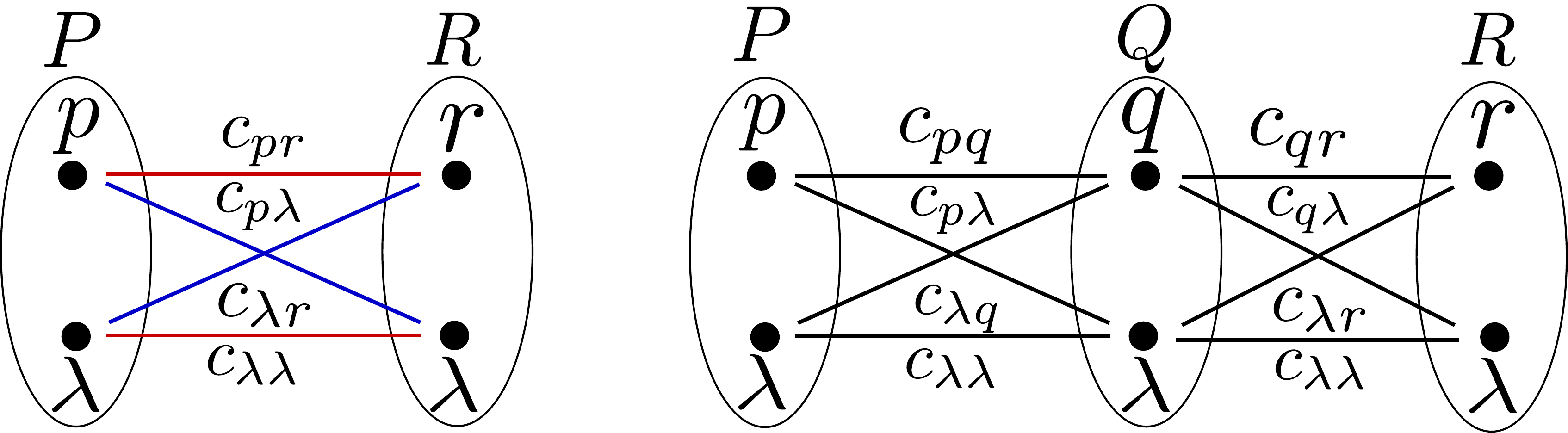}
\caption{The cost function satisfies triangle inequality. (left)~The blue or the red matching may be the minimum weight maximum matching that corresponds to the cost of the edit operation. (right)~The matching between $P$ and $R$ is a composition of matching between $P, Q$ and $Q,R$. The inequality can be proved via case analysis by considering all possible compositions.}
\label{fig:proof}
\vspace{-0.15in}
\end{figure}

Consider three multisets as shown in Figure~\ref{fig:proof}. Using the above construction, we prove triangle inequality by considering the two cases, namely when the minimum weight matching is equal to either the red or blue matching.

\noindent\textbf{Case \textsc{red}}: $MM(P,R)$ is given by the red matching.
The cost of the relabel  $\gamma(p \longrightarrow r) = c_{pr} + c_{\lambda \lambda} = c_{pr}$.  Two different paths from $p$ lead to $r$, $p \longrightarrow q \longrightarrow r$ and $p \longrightarrow \lambda \longrightarrow r$. 
Consider the first path, 
\begin{align}
\gamma(p \longrightarrow r) = c_{pr}  = L_\infty(p,r) &\le L_\infty(p,q) + L_\infty(q,r)\\ &= c_{pq} + c_{qr}\\ &= \gamma(p \longrightarrow q) + \gamma(q \longrightarrow r)
\end{align}
Now, let us consider the second path. If $c_{pr} \le c_{p\lambda} + c_{\lambda r}$ then  $c_{pr} \le c_{p\lambda} + c_{\lambda q} + c_{q\lambda} + c_{\lambda r}$  and we are done. Else, we have two sub-cases 
\begin{align}
c_{pr} &> c_{p\lambda} + c_{\lambda q} + c_{q\lambda} + c_{\lambda r} > c_{p\lambda} + c_{\lambda r}, \mbox{ or}\\ c_{pr} &> c_{p\lambda} + c_{\lambda r} \mbox{ but } c_{pr} < c_{p\lambda} + c_{\lambda q} + c_{q\lambda} + c_{\lambda r}. 
\end{align}
In both sub-cases, we have a matching with weight $MM'(P,R) = c_{p\lambda} + c_{\lambda r} \le MM(P,R) = c_{pr} + c_{\lambda \lambda}$, which contradicts our assumption.

The cost $\gamma(\lambda \longrightarrow \lambda) = c_{\lambda \lambda} = 0$. Both paths via $Q$, $\lambda \longrightarrow \lambda \longrightarrow \lambda$ and $\lambda \longrightarrow q \longrightarrow \lambda$, should necessarily have a non-zero total cost. So, the inequality holds trivially.

\noindent\textbf{Case \textsc{blue}}: $MM(P,R)$ is given by the blue matching.
The cost of the relabel is equal to the sum $ c_{p\lambda} + c_{\lambda r}$. We consider the two weights individually.

Two paths from $p \in P$ lead to $\lambda \in R$ via a node in $Q$, $p \longrightarrow \lambda \longrightarrow \lambda$ and $p \longrightarrow q \longrightarrow \lambda$. Similarly,  two paths from $\lambda \in P$ lead to $r \in R$, $\lambda \longrightarrow \lambda \longrightarrow r$ and $\lambda \longrightarrow q \longrightarrow r$.

In both cases, triangle inequality holds trivially for the first path via $Q$, $c_{p \lambda} \le c_{p  \lambda} + c_{\lambda \lambda}$ and $c_{\lambda r} \le c_{\lambda \lambda} + c_{\lambda r}$. We need to show that the inequality holds for the second paths as well. 
From the persistence diagram, we observe that $L_\infty(p, \lambda) \le L_\infty(p,q) + L_\infty(q, \lambda)$ for all $q$, even when $q$ lies on the perpendicular from $p$ onto the diagonal. So, $c_{p\lambda} \le c_{pq} + c_{q\lambda}$. A similar argument can be used to show that $c_{\lambda r} \le c_{\lambda q} + c_{qr}$.

The red and blue cases together imply that the cost $C_{W}$ satisfies the triangle inequality and is therefore a metric. It follows that the tree edit distance measure $D$ is also a metric. 

\subsection{Handling instabilities}\label{hi}
\begin{figure}
\centering
\includegraphics[width=0.30\textwidth]{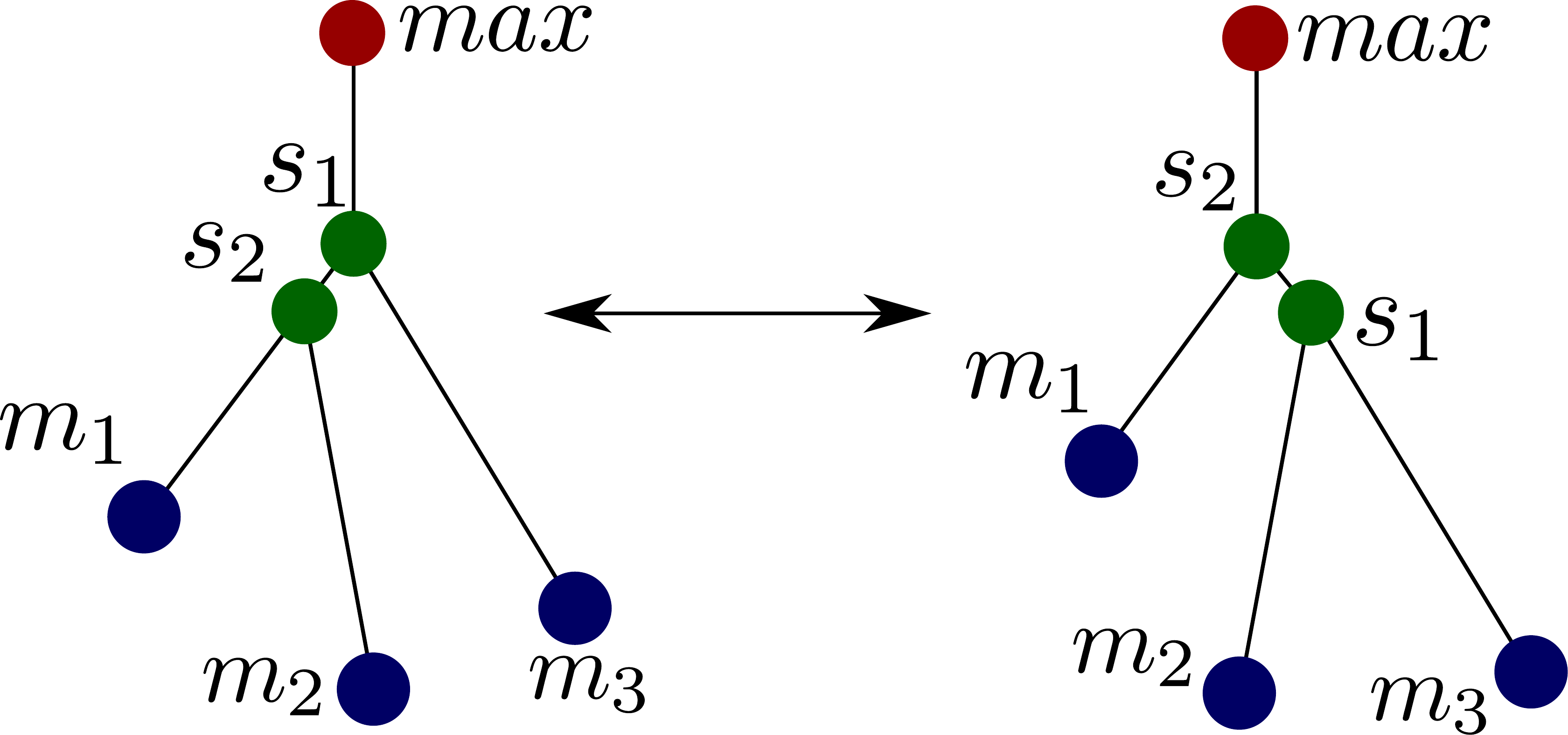}
\caption{Illustrating instabilities. Since the difference in the  function values between $s_1$ and $s_2$ is small, a slight perturbation leads to a change in the structure of the tree, which affects the distance measure.}
\label{fig:instab}
\vspace{-0.15in}
\end{figure}

Saikia et al.~\cite{Saikia2014} discuss two kinds of instabilities, \emph{vertical} and \emph{horizontal}, that affect branch decompositions and hence the distance measures. Figure~\ref{fig:instab} illustrates how horizontal instability can occur. In our case, the horizontal stability has a more drastic effect on the measure because
\begin{itemize}
\item It changes the persistence pairing, which in turn affects the cost.
\item It also changes the subtrees thereby affecting the matching found by the algorithm.
\end{itemize}
We employ a  strategy similar to the one used for branch decompositions by Thomas and Natarajan~\cite{thomas2011} and apply it to merge trees. We introduce a stability parameter $\varepsilon$ and use it to determine how  to merge simple saddles into a multi-saddle where instabilities occur. We merge the saddles in a bottom-up manner as follows. Begin from the lower saddle $s_l$ that is further from the root and merge it into a higher saddle  $s_h$ that is nearer to the root if the function difference $ |f(s_h)-f(s_l)| < \varepsilon$. Repeat this process until none of the saddles satisfy the merging condition. In the implementation, the multi-saddle is represented by the saddle with the highest persistence. We compute the distance between the stabilized trees. Optionally, a fixed value may be added to the final distance to incorporate the cost incurred due to the stabilization. In Section~\ref{sec:stb}, we experimentally analyze how varying the stability parameter $\varepsilon$ affects the distance measure.

\subsection{Algorithm}\label{algo}
We adapt Zhang's algorithm~\cite{Zhang1996} (See supplementary material, Section 2) with the edit costs discussed in Section~\ref{cm} to compute the tree edit distance between merge trees. The input to this algorithm is a pair of merge trees that are stabilized using the strategy described in Section~\ref{hi}.

\subsection{Implementation}
The computation proceeds in a bottom up manner. Distances for the subtrees are computed and stored in a table. These are next used for computing distances between subtrees at higher levels of the merge trees. This proof of concept implementation does not include code and memory optimizations for efficiently computing and storing the dynamic programming tables. We use the simple Kuhn-Munkres algorithm~\cite{kuhn1955} for computing $MM(i,j)$. We still observe reasonable running times for most of the data sets as reported in the individual experiments in the following section.

\section{Experiments and case studies}\label{sec:ExpCaseStudies}
We demonstrate the utility of the tree edit distance measure by applying it to analyze time-varying data, to study symmetry in scalar fields, for summarizing data, and for shape matching. We use the Recon library~\cite{doraiswamy2013} to compute merge trees, the algorithm described in Section~\ref{algo} to compute the tree edit distance between the merge trees, and Paraview~\cite{ahrens2005} together with the Topology ToolKit TTK~\cite{tierny2018} to generate  renderings of the merge trees together with the scalar fields. We uniformly use the $L_{\infty}$ cost $C_W$ (\ref{cw}). All experiments were performed on a machine with an Intel Xeon CPU with $8$ cores running at $2.0$ GHz and $16$ GB main memory. 

\subsection{Understanding the distance measure}

\begin{figure}
\centering
\subfigure[Three scalar field $f_1,f_2,f_3$]{\includegraphics[width=0.165\textwidth,angle=270,origin=c]{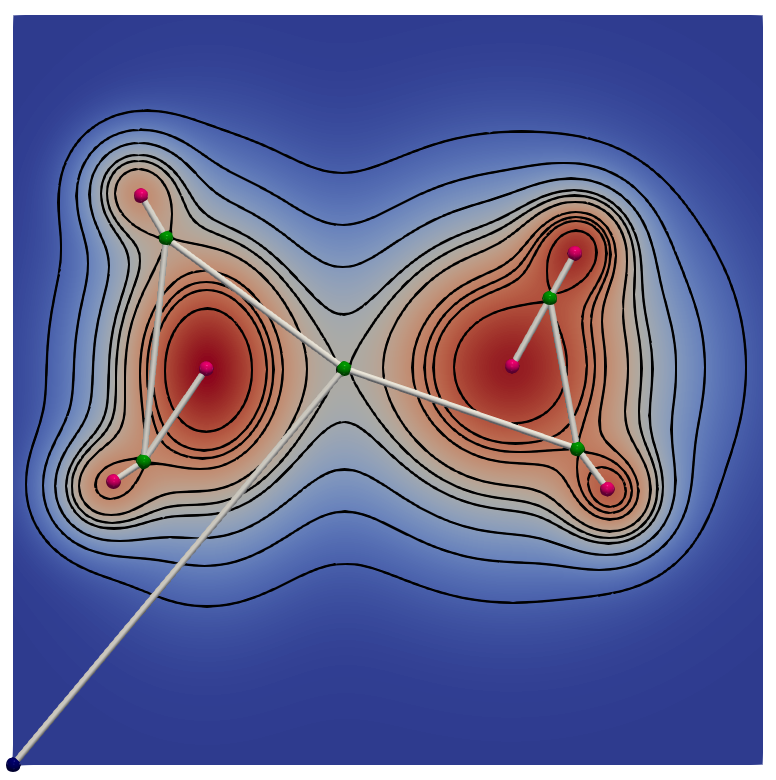}\includegraphics[width=0.165\textwidth,angle=270,origin=c]{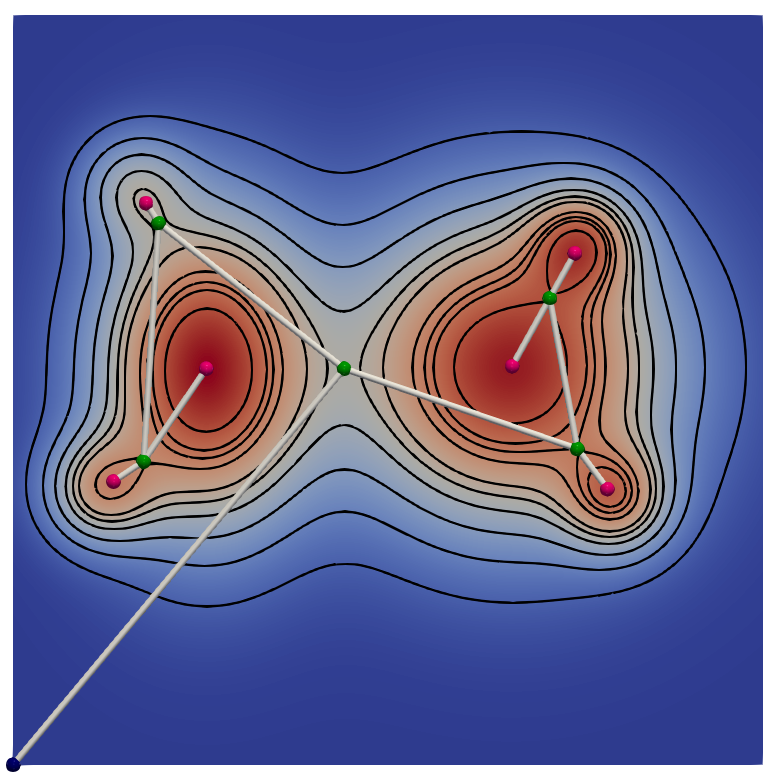}\includegraphics[width=0.165\textwidth,angle=270,origin=c]{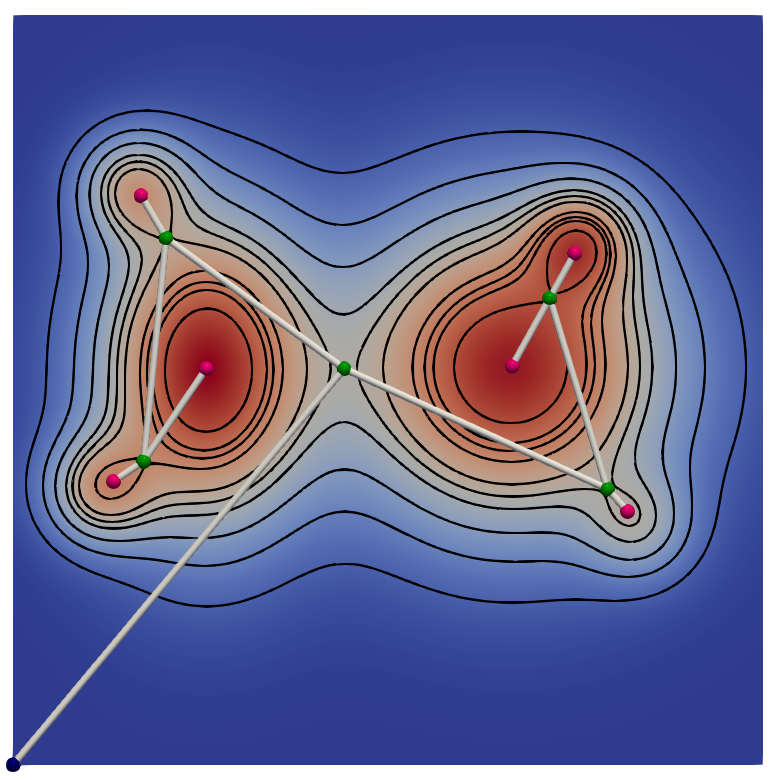}
\label{fig:u1}}

\subfigure[Merge tree driven segmentation for each field.]{\includegraphics[width=0.165\textwidth,angle=270,origin=c]{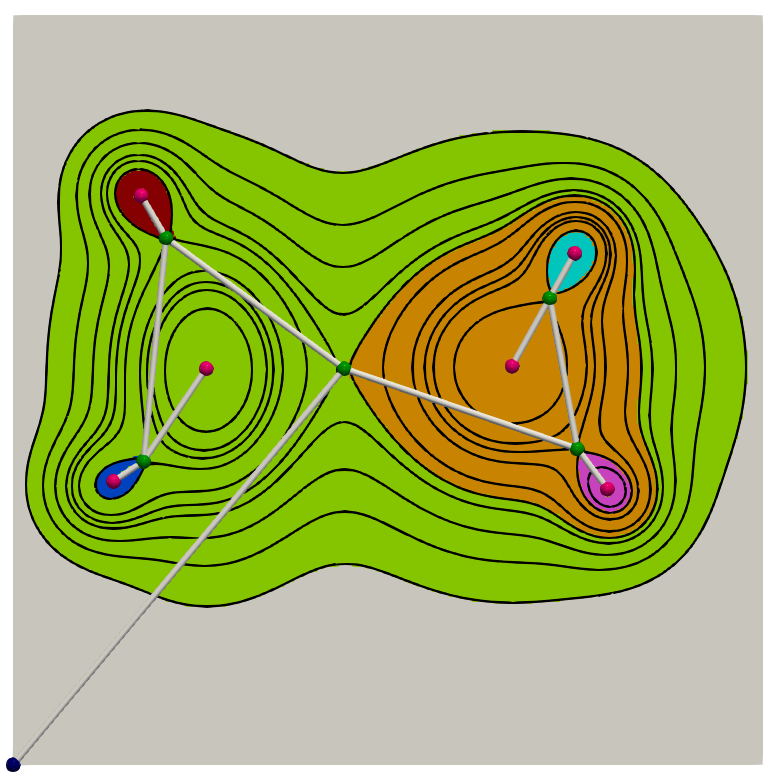}\includegraphics[width=0.165\textwidth,angle=270,origin=c]{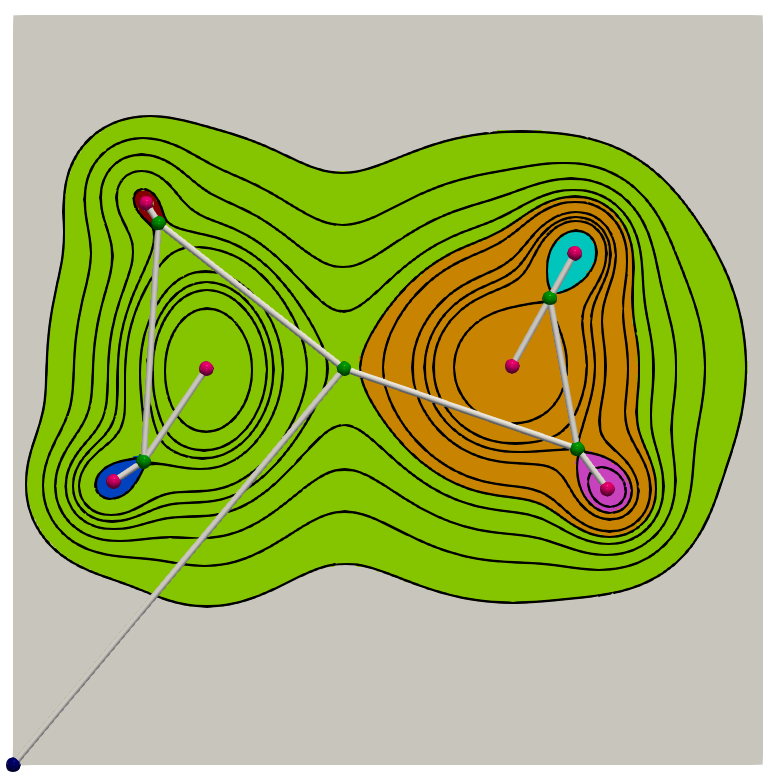}\includegraphics[width=0.165\textwidth,angle=270,origin=c]{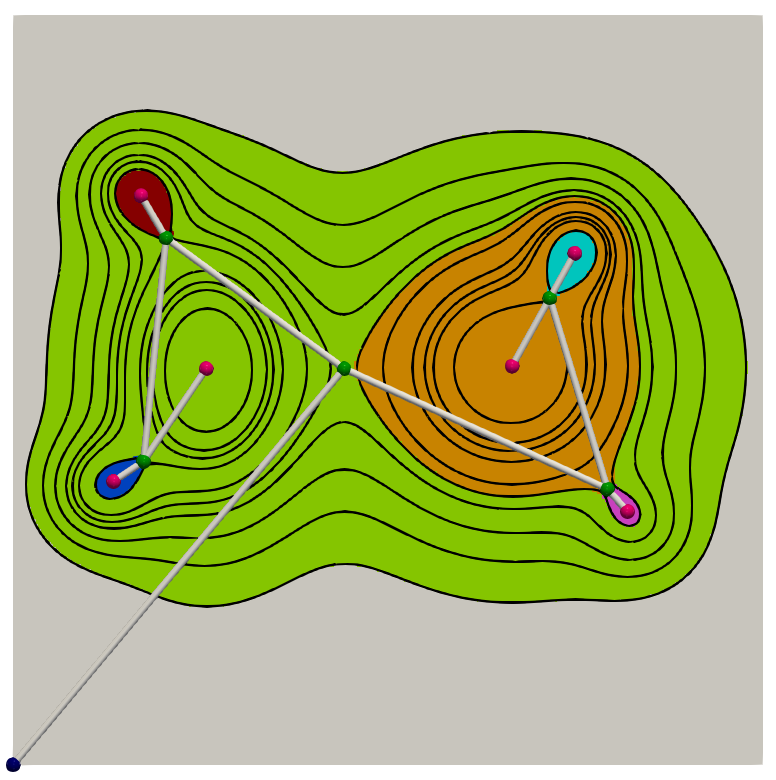}
\label{fig:u4}}

\subfigure[Mapping determined by $W_1$ for ($f_1, f_2$) and ($f_2, f_3$)]{\includegraphics[width=0.2\textwidth]{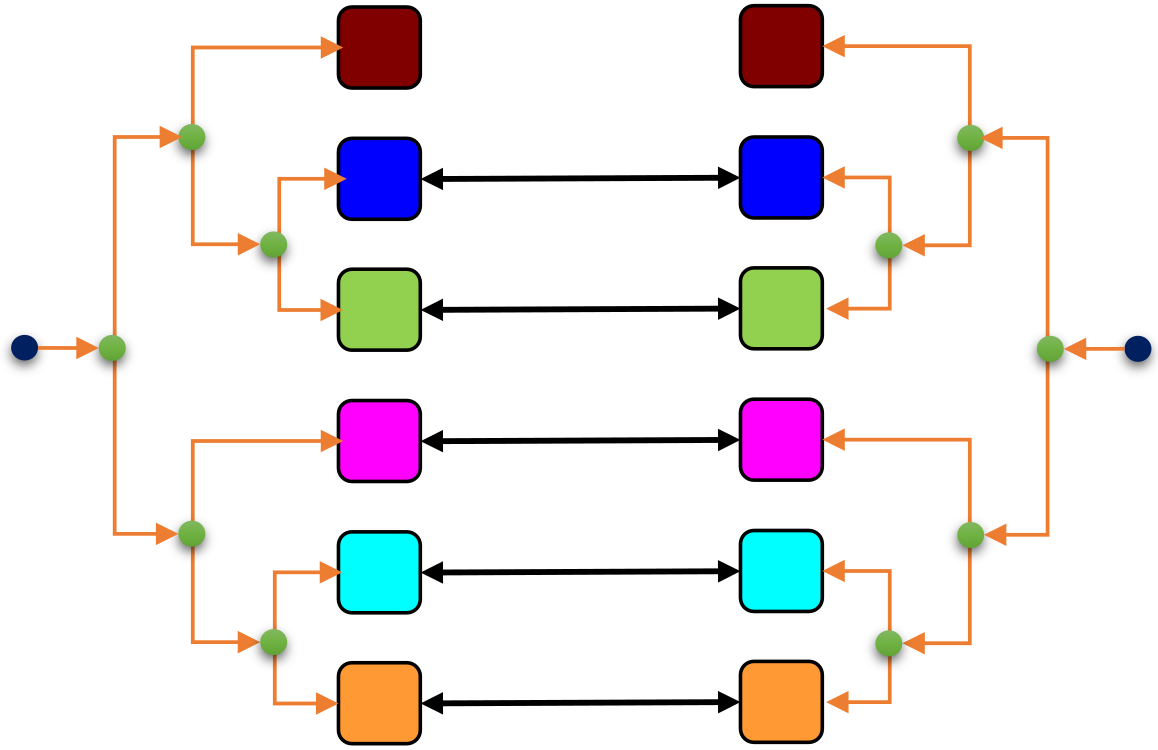}~\includegraphics[width=0.2\textwidth]{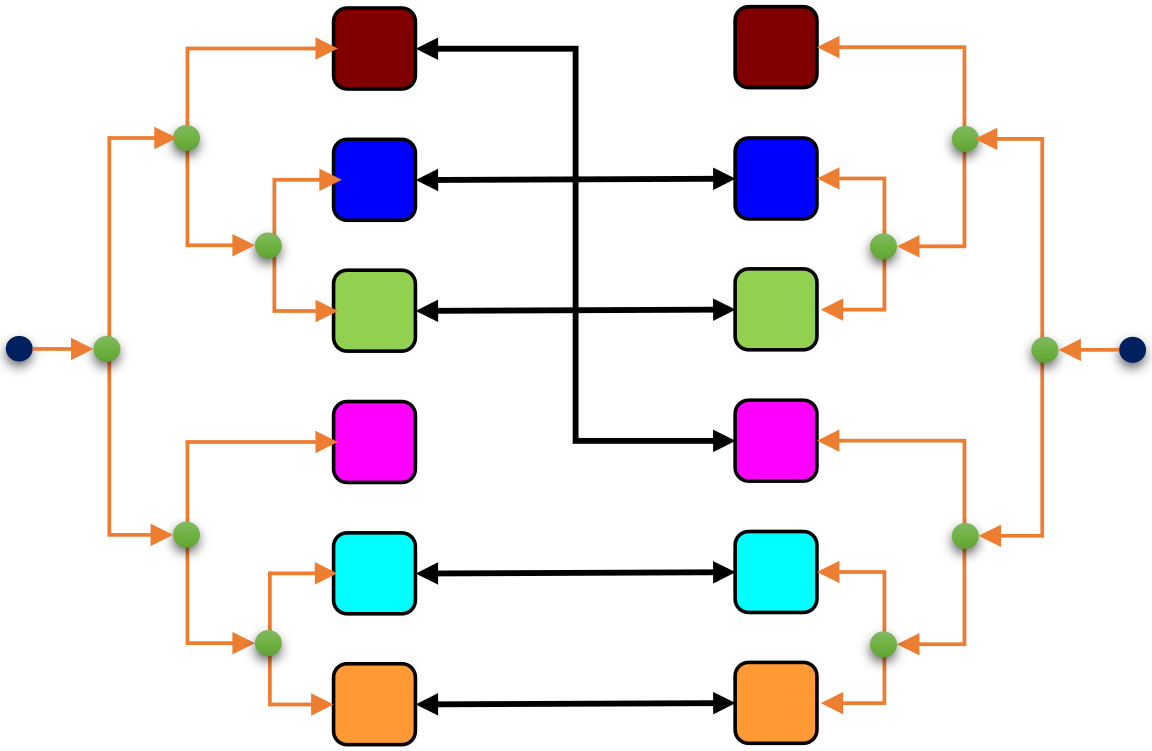}
\label{fig:u9}}

\subfigure[Mapping determined by $D$ for ($f_1, f_2$) and ($f_2, f_3$)]{\includegraphics[width=0.2\textwidth]{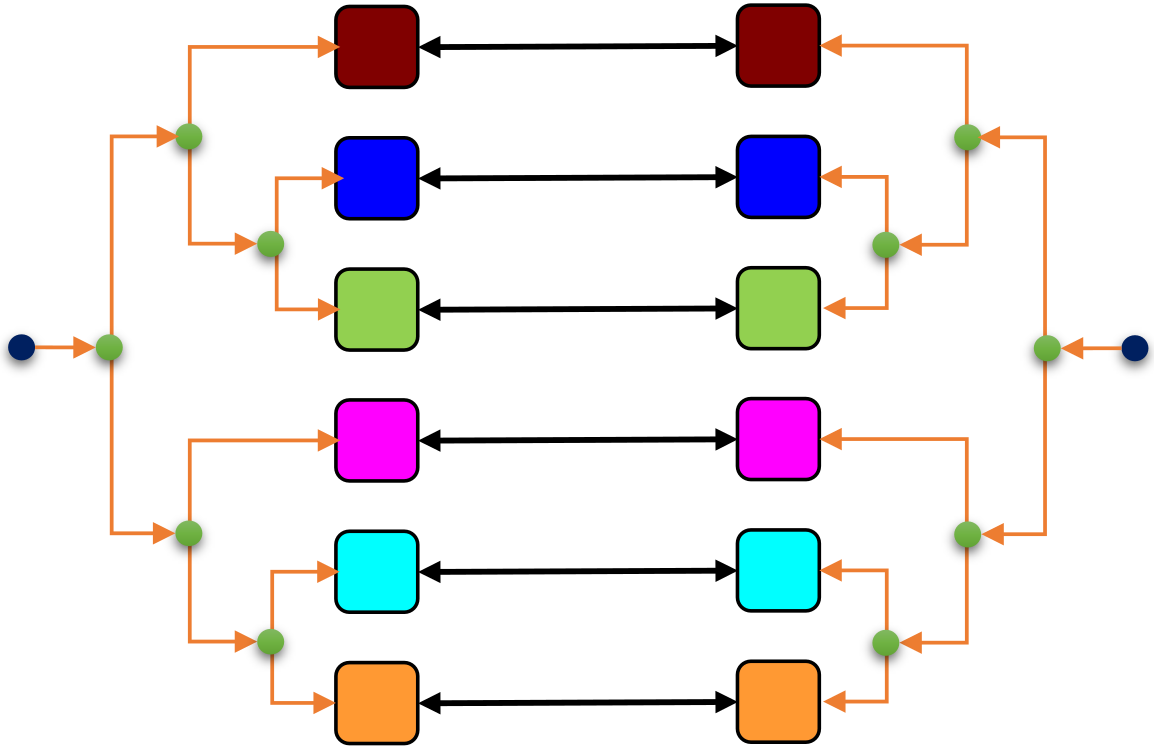}~\includegraphics[width=0.2\textwidth]{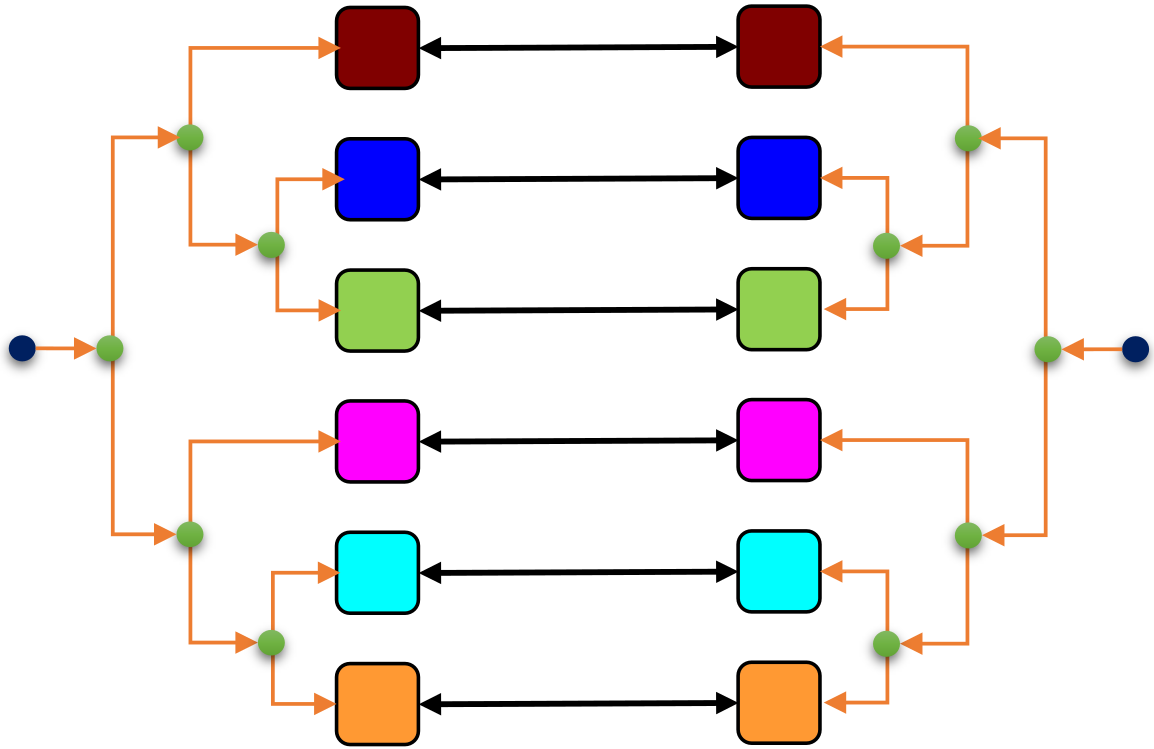}
\label{fig:u7}}

\caption{Comparing mappings established by tree edit distance measure $D$ and Wasserstein distance $W_1$. \subref{fig:u1}~Three scalar functions $f_1, f_2, f_3$ in the synthetic data set. \subref{fig:u4}~Regions corresponding to the maxima and arcs incident on them in the merge trees of $f_1, f_2, f_3$. Each region is assigned a unique color.  \subref{fig:u9}~Mapping determined by $W_1$ between ($f_1, f_2$) and between ($f_2, f_3$). \subref{fig:u7}~Mapping determined by the tree edit distance $D$ between ($f_1, f_2$) and between ($f_2, f_3$). Merge tree nodes and their corresponding spatial regions have the same color.
}
\label{fig:ud}
\vspace{-0.15in}
\end{figure}

We construct three synthetic datasets to understand the difference between the tree edit distance $D$ and other well known distances between topological structures. The scalar functions $f_1, f_2, f_3$ are sums of gaussians whose extrema are fixed in space. The scalar values change in a controlled manner for the three functions so that the values at the extrema increase / decrease monotonically as we step from $f_1$ to  $f_2$ to $f_3$. We compute the tree edit distances together with the corresponding mapping for each pair $(f_1, f_2)$ and  $(f_2, f_3)$.

Figure~\ref{fig:ud} shows the three scalar functions $f_1,f_2,f_3$. We observe in Figure~\ref{fig:u7} that $D$ establishes intuitively correct mappings. The mappings also preserve the tree hierarchy. On the other hand, the Wasserstein distance $W_1$  (Figure~\ref{fig:u9}, left) maps the brown regions to the null character. The tree edit distance prefers the relabel over a sequence of delete-insert operations. The reason $W_1$ does not find the correspondence between the two brown nodes is because their birth-death intervals do not overlap. As a result, these nodes are mapped to the null character \emph{i.e.}, inserted or deleted. The intervals corresponding to the two nodes in question are $[1.17,1.39]$ in $f_1$ and $[1.06, 1.08]$ in $f_2$.

We also see from Figure~\ref{fig:u9} that $W_1$ maps the brown region to the magenta region, thereby mapping nodes that lie within different subtrees. This also causes a pair of nodes being mapped to the null character. The tree edit distance $D$ is constrained to map disjoint subtrees to disjoint subtrees and establishes a better mapping. To summarize, $D$ in general establishes mappings that are better than $D_B$ and $W_1$ because it is aware of the structure of the merge tree and preserves the hierarchy captured in the tree. 

\subsection{Comparison with other distance measures}\label{sec:stb}

\begin {figure}
\centering
\begin{tikzpicture}
\begin{axis}[name=plot1, width=0.48\textwidth,
    height=0.13\textheight,
    xmin=0, xmax=220,
    ymin=0, ymax=1.2,
    tick pos=left,
    tick align=outside,
    legend columns=3,
    ymajorgrids=false,
    tick label style={font=\boldmath},
]
 
\addplot[line width=0.75pt,color=blue]
    table[x=t,y=d0,col sep=comma]{plot.csv};
    
\addplot[line width=0.75pt,color=teal]
    table[x=t,y=w,col sep=comma]{plot.csv};

\addplot[line width=0.75pt,color=magenta]
    table[x=t,y=b,col sep=comma]{plot.csv};
    
    \legend{TED $D$,$W_1$,$D_B$}
\end{axis}

\begin{axis}[name=plot2,at={($(plot1.south)-(0,1.45cm)$)},anchor=north, width=0.48\textwidth,
    height=0.13\textheight,
    xmin=0, xmax=220,
    ymin=0, ymax=1.2,
    tick pos=left,
    tick align=outside,
    legend columns=3,
    ymajorgrids=false,
    tick label style={font=\boldmath},
]
 
\addplot[line width=0.75pt,color=blue]
    table[x=t,y=d0,col sep=comma]{plot.csv};
    
\addplot[line width=0.75pt,color=green]
    table[x=t,y=d1.5,col sep=comma]{plot.csv};    
    
\addplot[line width=0.75pt,color=red]
    table[x=t,y=d5,col sep=comma]{plot.csv};
    
\legend{$\varepsilon$ = 0\%,$\varepsilon$ = 1.5\%,$\varepsilon$ = 5\%}

\end{axis}

\begin{axis}[name=plot3,at={($(plot2.south)-(0,1.45cm)$)},anchor=north,width=0.48\textwidth,
    height=0.13\textheight,
    xmin=0, xmax=220,
    ymin=0, ymax=1.2,
    tick pos=left,
    tick align=outside,
    legend columns=3,
    ymajorgrids=false,
    tick label style={font=\boldmath},
    ] 

\addplot[line width=0.75pt,color=red]
    table[x=t,y=d5,col sep=comma]{plot.csv};
    
\addplot[line width=0.75pt,color=cyan]
    table[x=t,y=d10,col sep=comma]{plot.csv};
    
\addplot[line width=0.75pt,color=brown]
     table[x=t,y=d20,col sep=comma]{plot.csv};

\legend{$\varepsilon$ = 5\%,$\varepsilon$ = 10\%,$\varepsilon$ = 20\%},
\end{axis}

\begin{axis}[name=plot4,at={($(plot3.south)-(0,1.75cm)$)},anchor=north,width=0.48\textwidth,
    height=0.13\textheight,
    xlabel={Timesteps},
    xmin=0, xmax=220,
    ymin=0, ymax=1.2,
    tick pos=left,
    tick align=outside,
    legend columns=2,
    ymajorgrids=false,
    tick label style={font=\boldmath},
    ] 

\addplot[line width=0.75pt,color=brown]
     table[x=t,y=d20,col sep=comma]{plot.csv};
    
\addplot[line width=0.75pt,color=yellow]
    table[x=t,y=d50,col sep=comma]{plot.csv};

\addplot[line width=0.75pt,color=violet]
    table[x=t,y=d100,col sep=comma]{plot.csv};
    
\addplot[line width=0.75pt,color=teal]
    table[x=t,y=w,col sep=comma]{plot.csv};

\legend{$\varepsilon$ = 20\%,$\varepsilon$ = 50\%,$\varepsilon$ = 100\%,$W_1$},
\end{axis}
\end{tikzpicture}

\caption{Comparing distance measures on the von K\'arm\'an vortex street dataset. (top)~Plot of distance measures between the first time step and others when stability parameter $\varepsilon$ is set to $0$ and comparison with the Wasserstein distance $W_1$ and bottleneck distance $D_B$.  (rows 2-4)~~Effect of stabilization parameter $\varepsilon = 0,1.5,5,10,20,50,100\%$  and comparison with Wasserstein distance $W_1$. Results are shown in three plots to reduce clutter.}
\label{fig:graphs}
\vspace{-0.25in}
\end{figure}
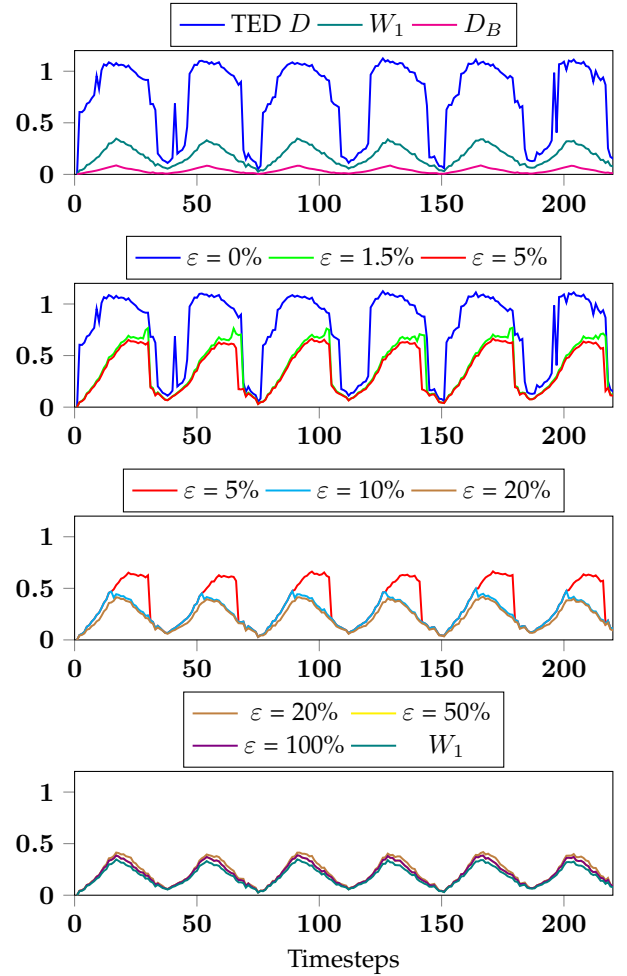

We compare the proposed tree edit distance measure $D$ with existing measures such as bottleneck distance $D_B$ and Wasserstein distance $W_1$ via computational experiments on the 2D B\'enard-von K\'arm\'an vortex street  dataset~\cite{weinkauf2010}. Figure~\ref{fig:timevary} shows a few time steps of the data, which represents flow around a cylinder. The dataset contains the velocity magnitude on a $400 \times 50$ grid over $1001$ time steps. Each split tree contains approximately $55-65$ nodes. We calculate $D$ and plot it together with the bottleneck and Wasserstein distance, see top row of Figure~\ref{fig:graphs}. The tree edit distance $D$ is always greater than $W_1$ and $D_B$. Indeed, $D$ is likely to be more discriminative than $W_1$ and $D_B$ because it incorporates the structure of the merge tree in addition to the persistence pairs. 

We also compute and plot $D$ for increasing values of the stability parameter $\varepsilon$. The values of $\varepsilon$ are reported as a percentage of the maximum persistence of the particular dataset. While there are some anomalies for small values of $\varepsilon$, in general we observe in Figure~\ref{fig:graphs} that with increase in $\varepsilon$, $D$ tends towards $W_1$.  For a high enough value of $\varepsilon$, $D$ becomes almost equal to $W_1$. The reason for this behavior is that the bottleneck/Wasserstein distance does not consider the structure of the trees. Increasing the stability parameter transforms the tree to become more like a bush. Finally, all the nodes become children of the root thereby simplifying and eliminating the tree structure. Varying $\varepsilon$ from $0-5\%$ results in a decrease of up to $25$ nodes in the split tree. Further increasing $\varepsilon$ led to an additional reduction by   only $1-2$ nodes. We observe this trend in the distance plots also.

\subsection{Periodicity in time-varying data}

\begin{figure}
\centering
\subfigure[Three time steps from the flow around a cylinder simulation.]{\label{fig:timevary}\includegraphics[width=0.48\textwidth]{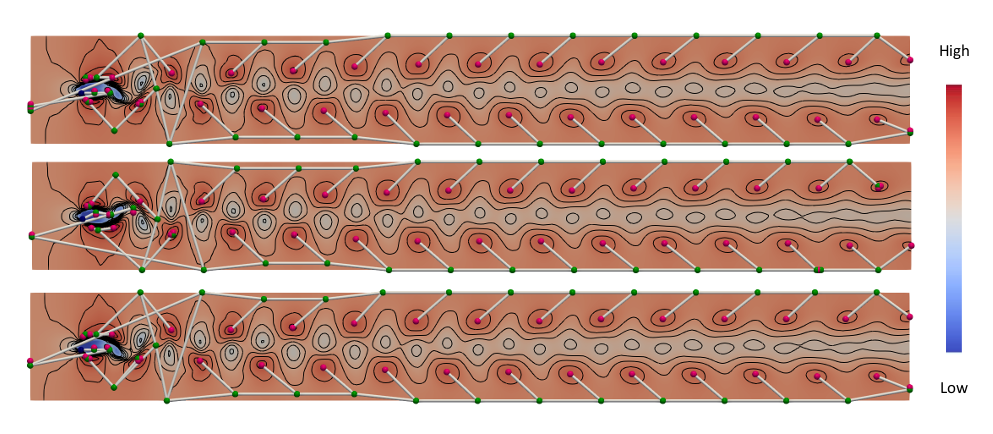}}
\subfigure[Distance matrix highlights the periodicity.]{\label{fig:tmat}\includegraphics[width=0.48\textwidth]{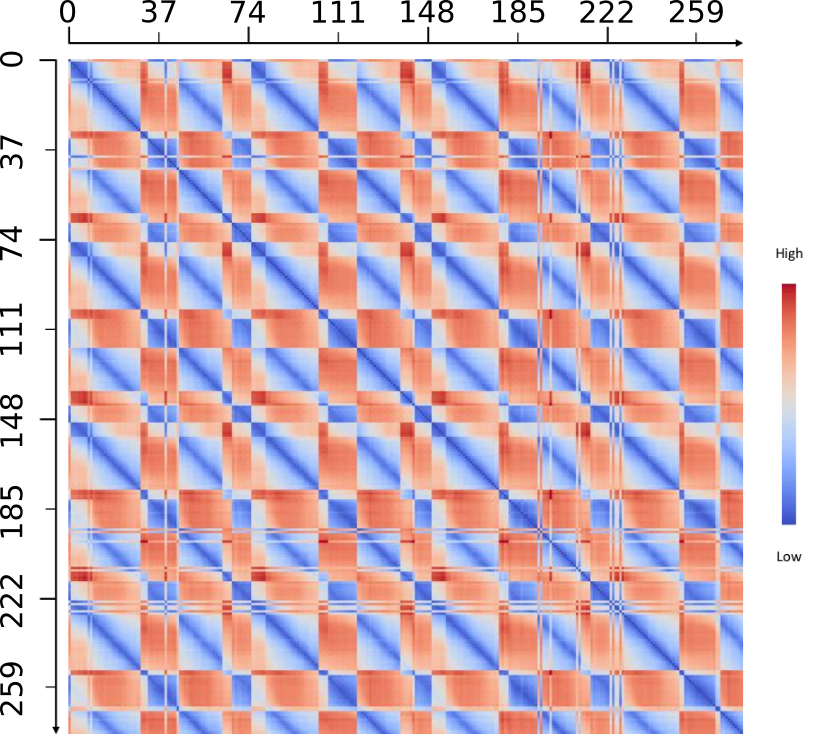}}
\caption{\subref{fig:timevary}~Time step 0 (top), 37 (middle) and 74 (bottom) of the von K\'arm\'an vortex street dataset.  The split tree and critical points are overlaid. \subref{fig:tmat}~A truncated version of the DM showing the tree edit distance measure between all pairs of time steps. Blue bands indicate periodicity with time period 74-75. A half period of 37, corresponding to the alternating nature of vortex shedding, is also visible.
}
\label{fig:period}
\vspace{-0.25in}
\end{figure}
Earlier studies of the B\'enard-von K\'arm\'an vortex street dataset have successfully identified periodicity in the dataset. Narayanan et al.~\cite{narayanan2015} detect both a half period of 38 and the full period of 75. We also aim to identify periodicity. Towards this, we compare the split tree of time step $1$ with the remaining $1000$ time steps of the dataset. We plot the tree edit distance for time steps $1-220$, see top plot of Figure~\ref{fig:graphs}. We rerun the experiment and compare all $1000$ time steps with all other time steps. The distances are stored in a distance matrix (DM). Each split tree contains approximately $55-65$ nodes. The distances were computed in parallel using 12 threads and took approximately $25$ minutes. A truncated version is shown in Figure~\ref{fig:tmat} for clarity. From Figure~\ref{fig:tmat}, we can also observe a periodicity of 37, which matches with the results reported by Narayanan et al.~\cite{narayanan2015}. The tree edit distance was computed in this experiment without stabilization.

\subsection{Topological effects of subsampling and smoothing}

\begin{figure}
\centering
\subfigure[$f_1$]{\includegraphics[width=0.14\textwidth]{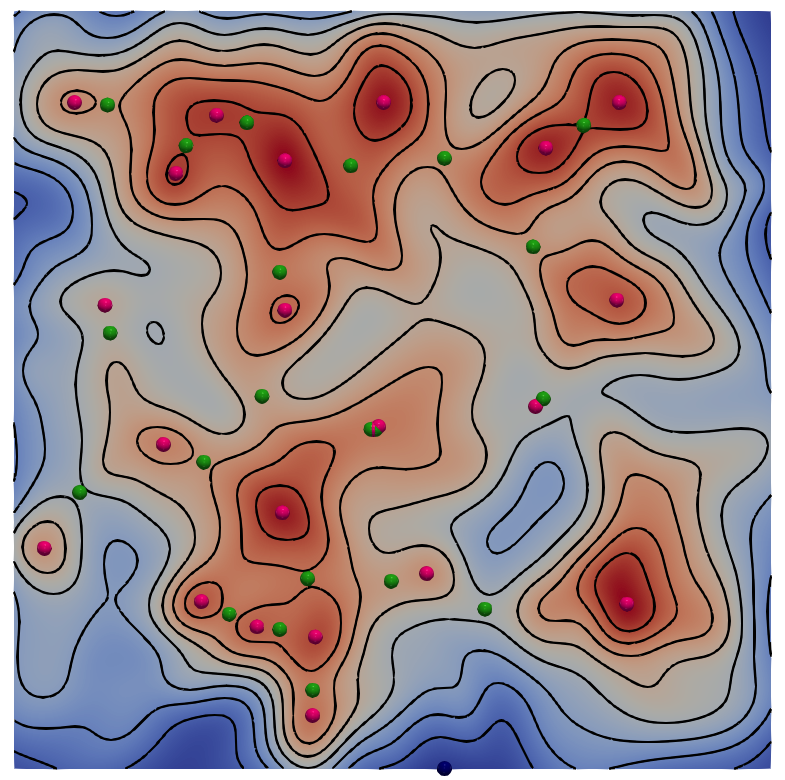}
\label{fig:ss1}}
~
\subfigure[subsampled $f_1$]{\includegraphics[width=0.14\textwidth]{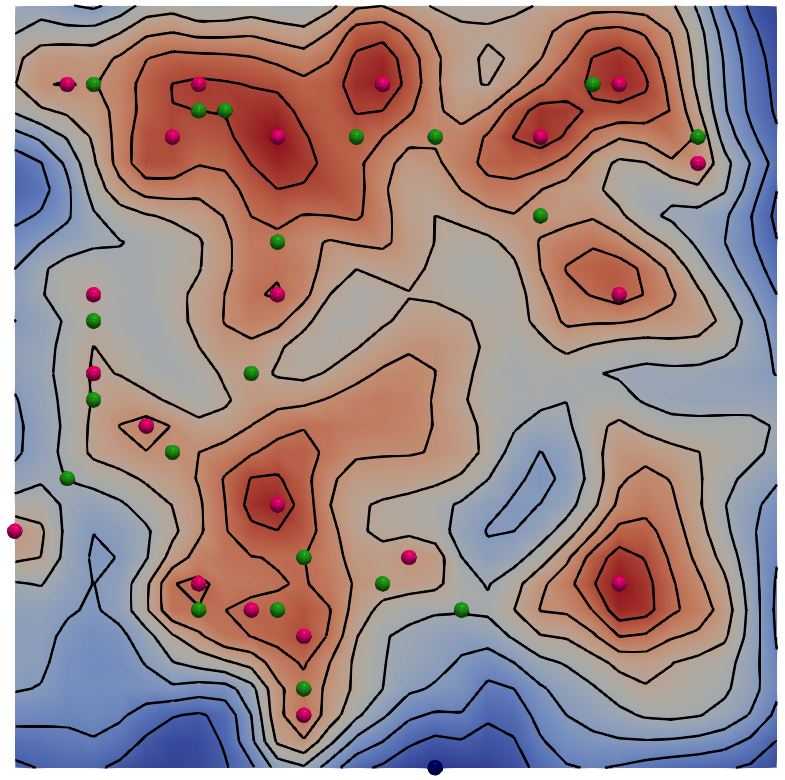}
\label{fig:ss2}}
~
\subfigure[smoothened $f_1$]{\includegraphics[width=0.14\textwidth]{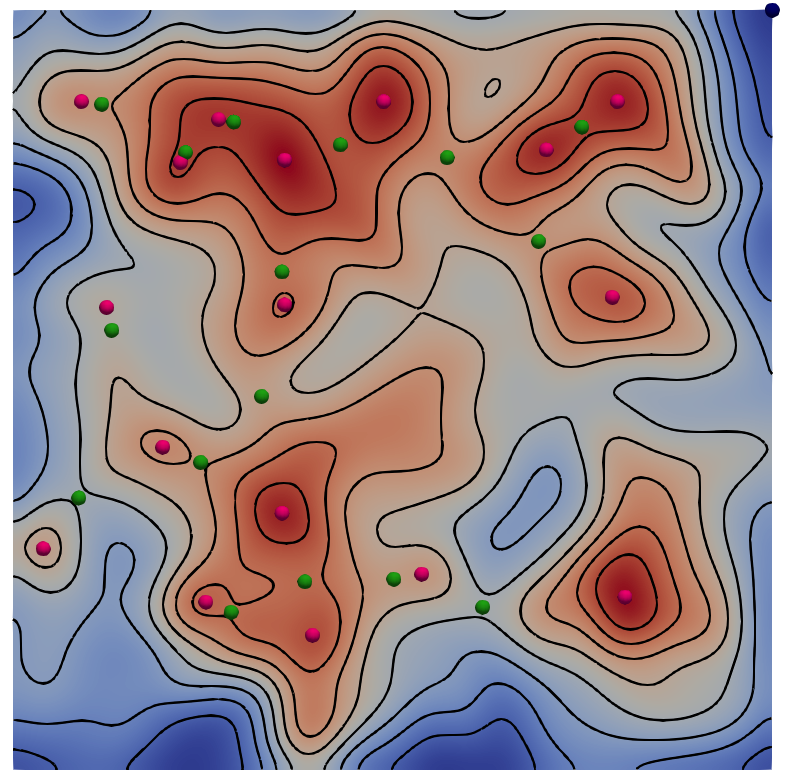}
\label{fig:ss3}}

\subfigure[$f_2$]{\includegraphics[width=0.14\textwidth]{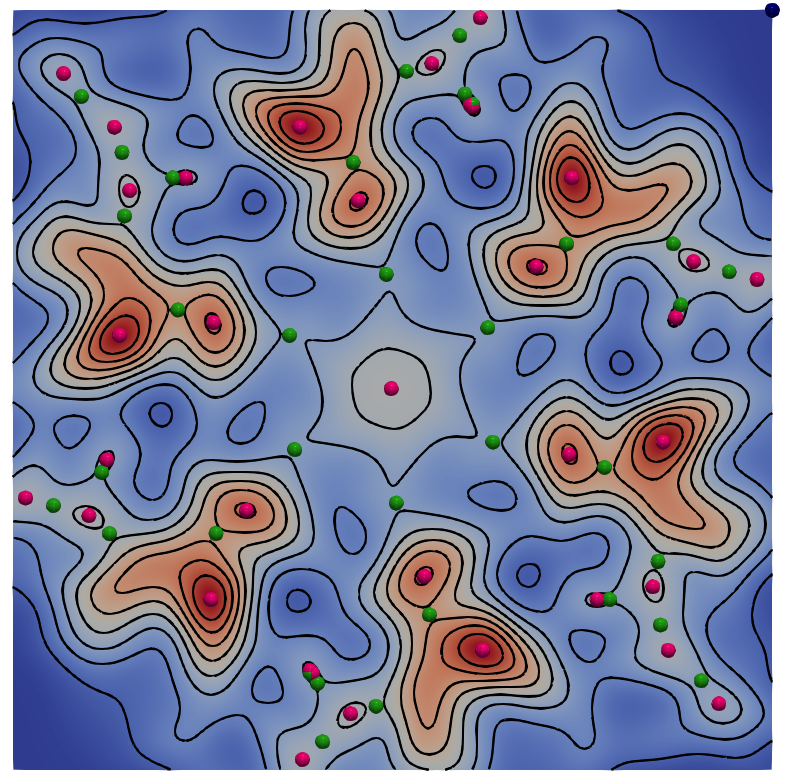}
\label{fig:ss4}}
~
\subfigure[subsampled $f_2$]{\includegraphics[width=0.14\textwidth]{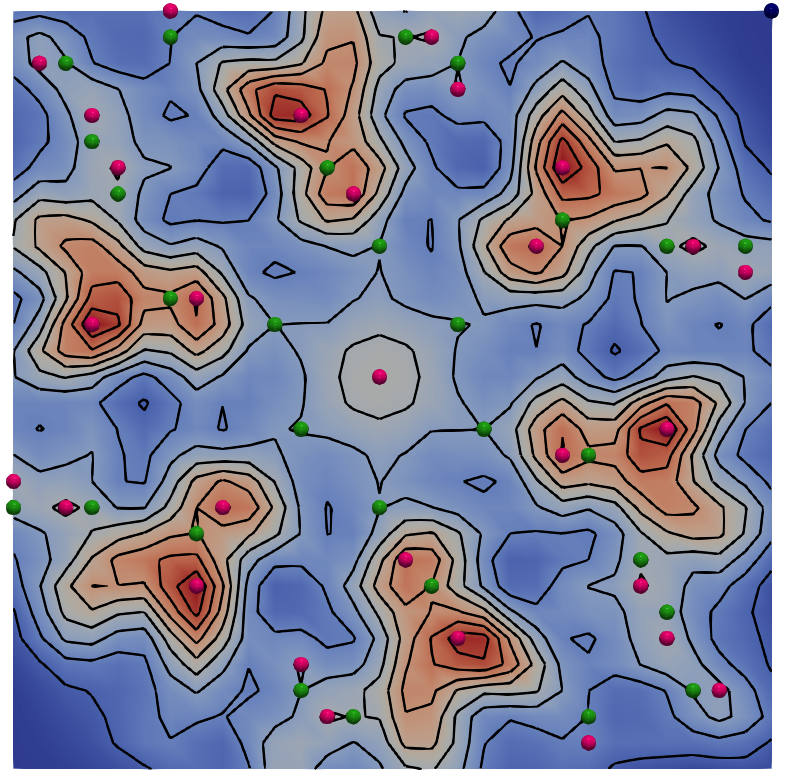}
\label{fig:ss5}}
~
\subfigure[smoothened $f_2$]{\includegraphics[width=0.14\textwidth]{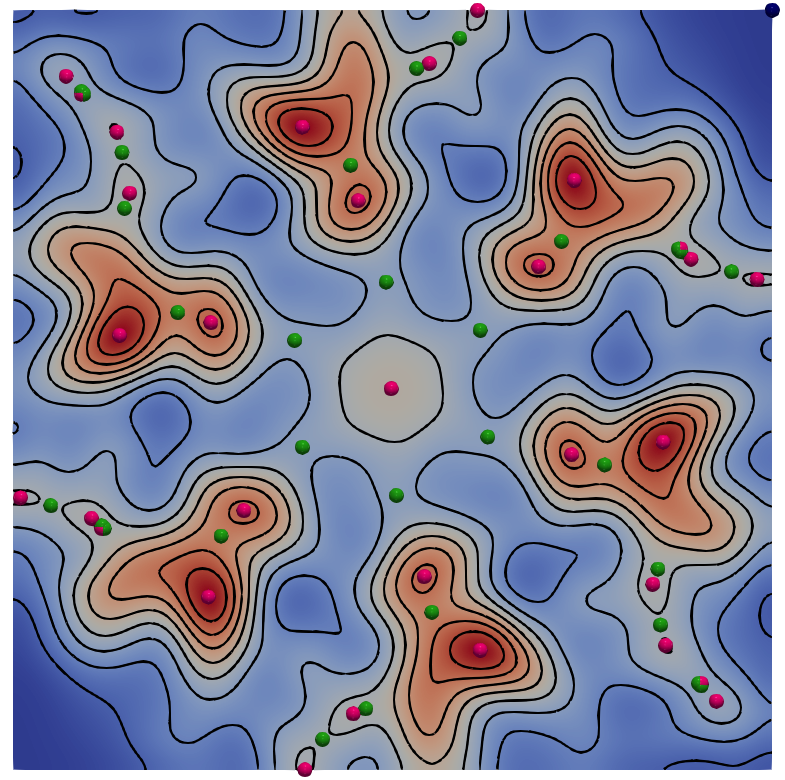}
\label{fig:ss6}}

\subfigure[DM for $f_1$, original and subsampled]{\includegraphics[width=0.14\textwidth]{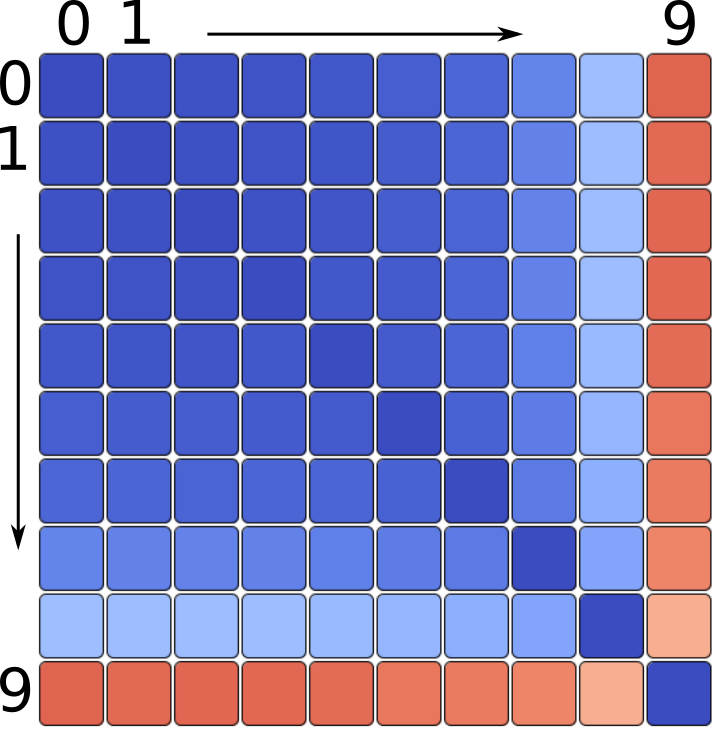}
\label{fig:sub1}}
~
\subfigure[DM for $f_2$, original and subsampled]{\includegraphics[width=0.14\textwidth]{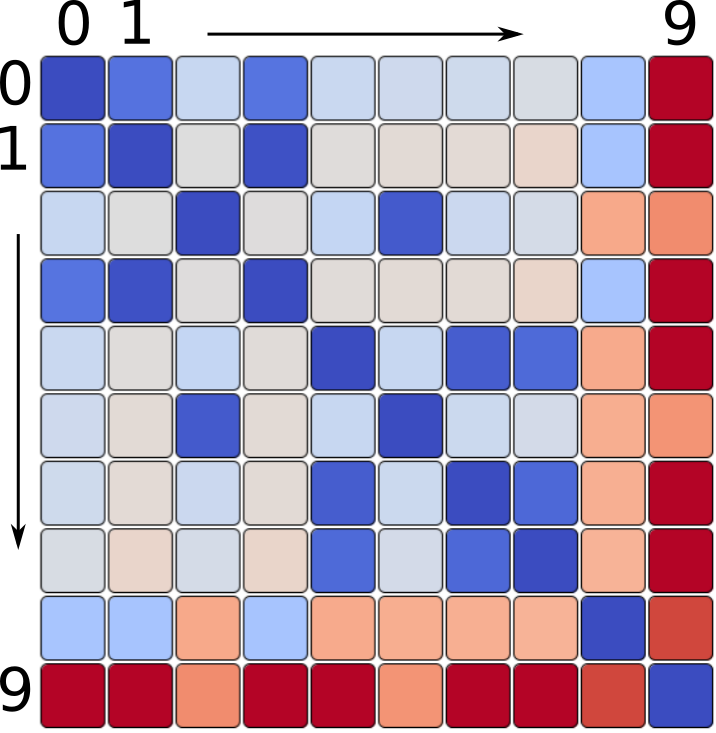}
\label{fig:sub2}}
~
\subfigure[DM for $f_2$, $\varepsilon = 0.5\%$]{\includegraphics[width=0.14\textwidth]{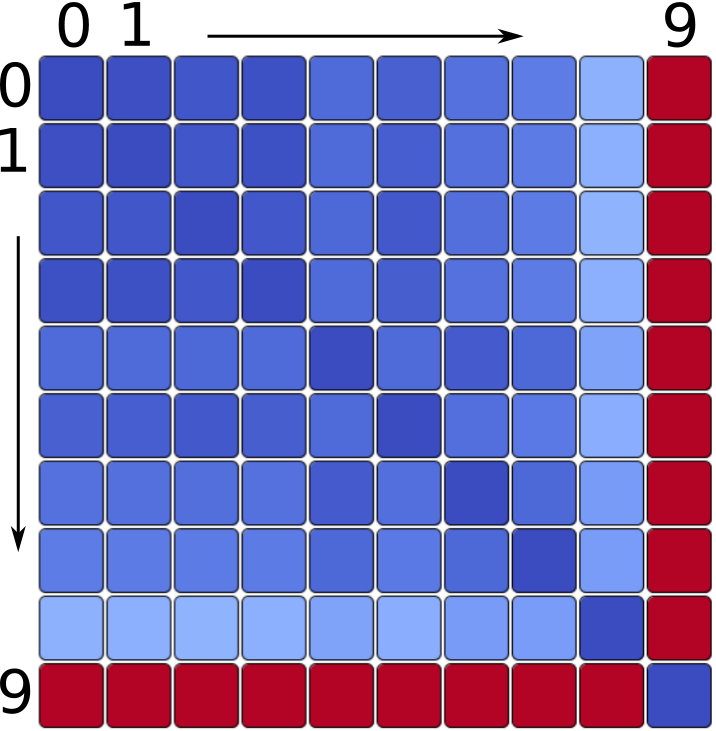}
\label{fig:sub3}}

\subfigure[DM for $f_1$, original and smoothened]{\includegraphics[width=0.14\textwidth]{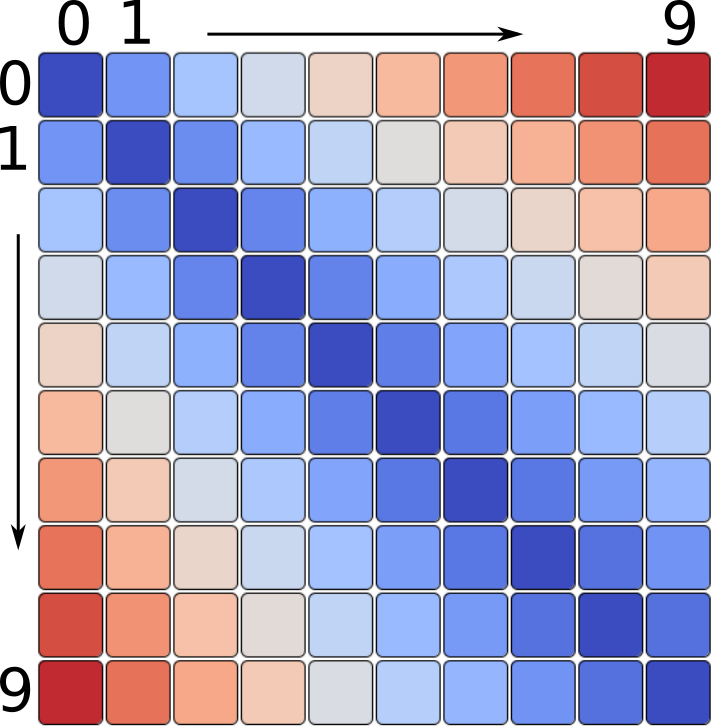}
\label{fig:smo1}}
~
\subfigure[DM for $f_2$, original and smoothened]{\includegraphics[width=0.14\textwidth]{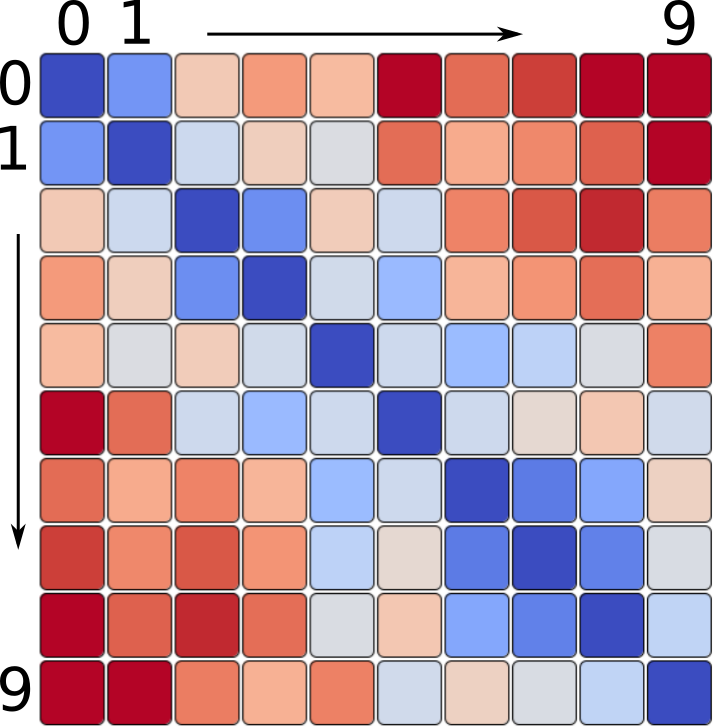}
\label{fig:smo2}}
~
\subfigure[DM for $f_2$, $\varepsilon = 0.5\%$]{\includegraphics[width=0.14\textwidth]{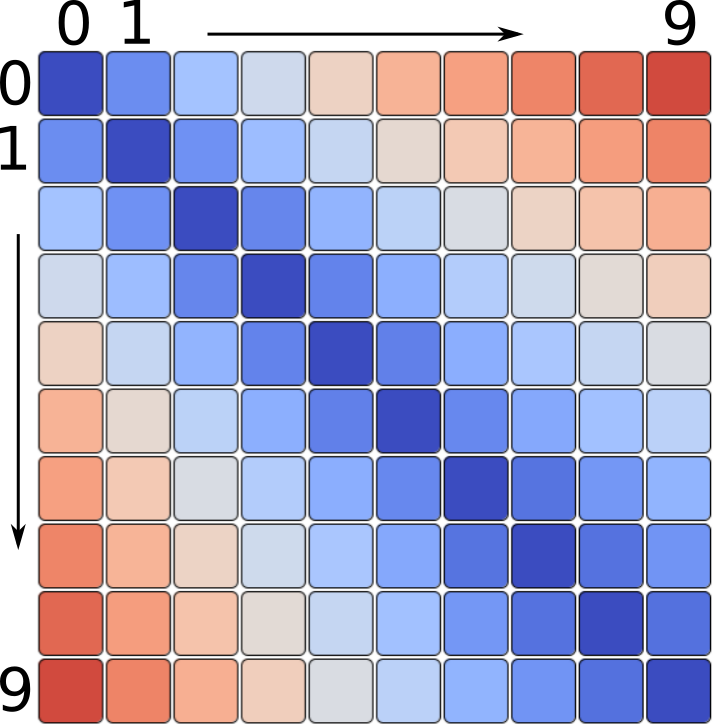}
\label{fig:smo3}}

\caption{Measuring the effect of subsampling and smoothing. \subref{fig:ss1},\subref{fig:ss4}~Two synthetic functions sampled over a $300 \times 300$ grid.
\subref{fig:ss2},\subref{fig:ss5}~Subsampled down to $30 \times 30$ over $9$ iterations.
\subref{fig:ss3},\subref{fig:ss6}~Smoothed in $9$ iterations.
\subref{fig:sub1}-\subref{fig:sub3}~DMs showing distance between all pairs of subsampled datasets without and with stabilization. \subref{fig:smo1}-\subref{fig:smo3}~DMs showing distances for all pairs of smoothed datasets. Row and column indices correspond to the iteration number, $0$ corresponds to the original, $9$ corresponds to the lowest resolution/extreme smoothing. Red indicates high and blue indicates low values. Colormaps for $f_1$ and $f_2$ are not on the same scale.
}
\label{fig:ss}
\vspace{-0.25in}
\end{figure}

The size of datasets are ever increasing and this mandates the use of subsampling and/or smoothing of the data as a preprocessing step. The aim of this preprocessing is to reduce the data size while ensuring a limited effect on geometric accuracy. However, the effect on the topological features of the scalar field is often not quantified. We want to observe how the tree edit distance measure captures these topological effects.

 We consider two synthetically generated datasets of size $300 \times 300$ (iteration $0$), see Figures~\ref{fig:ss1},~\ref{fig:ss4}. The data is downsampled over $9$ iterations to a $30 \times 30$ grid by reducing the number of samples in each dimension by $30$ within each iteration. We also apply $9$ iterations of laplacian smoothing on both $300 \times 300$  datasets. Next, we compare all merge trees corresponding to the subsampled and smoothed datasets pairwise. 

The distance matrix (DM) for the function $f_1$ indicates that the distances are monotonic, which conforms to the expected behavior. But we see a different pattern in the case of function $f_2$. A small stabilization applied on $f_2$ with $\varepsilon = 0.5\%$ results in distance matrices that conform to the expected behavior. This indicates that the stabilization may indeed be required, particularly when the scalar functions contain flat regions and multi-saddles. In both datasets, we notice that the distances between the lowest resolution ($30 \times 30$) dataset and others is relatively high. We identified two reasons for the high values. First, the number of critical points reduces significantly between iterations $8$ and $9$. For example, in the case of $f_2$, it goes down from $66-70$ in earlier iterations to $58$ in iteration $9$. Second,  the function value at the critical points in the lowest resolution dataset are also  different. Hence, the relabel costs increase significantly, up to a factor of 1.5 in some cases.

\subsection{Detecting symmetry / asymmetry}

\begin{figure}
\centering
\subfigure[Synthetic field]{\label{fig:sym1}\includegraphics[width=0.15\textwidth]{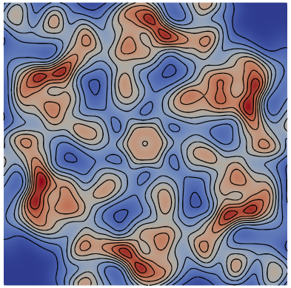}}
\subfigure[Segmentation]{\label{fig:sym2}\includegraphics[width=0.15\textwidth]{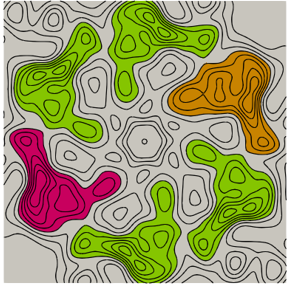}}
\subfigure[Distance matrix]{\label{fig:sym3}\includegraphics[width=0.15\textwidth]{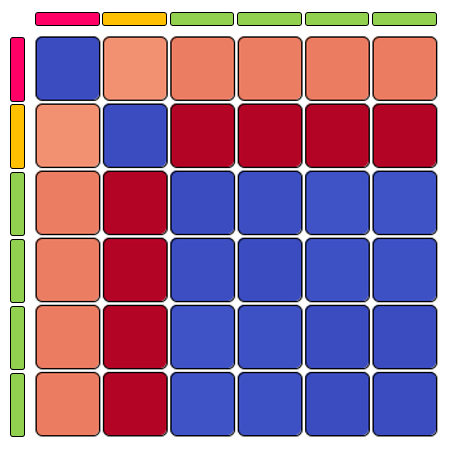}}
\caption{Identifying symmetry and asymmetry.  \subref{fig:sym1}~Sum of 2D gaussians. \subref{fig:sym2}~The DM indicates presence of a symmetric group containing 4 regions. Two regions are correctly identified as being different from the rest. \subref{fig:sym3}~DM between various subtrees of the merge tree.
}
\label{fig:symm}
\vspace{-0.15in}
\end{figure}

Identifying symmetric or repeating patterns in scalar fields enables feature-directed visualization. For example, it supports applications such as symmetry-aware transfer function design for volume rendering, anomaly detection, and query-driven exploration. A distance measure is central to any method for identifying symmetry. Consider the synthetic dataset in Figure~\ref{fig:symm} that contains six regions corresponding to six subtrees of the merge tree. Four regions colored green in Figure~\ref{fig:sym2} are symmetric copies. The remaining two regions, colored orange and magenta, are slightly perturbed to cause asymmetry. We compute the tree edit distance measure to compare each subtree corresponding to a region with other subtrees. The measure clearly distinguishes between symmetric and asymmetric regions as can be seen from the distance matrix (DM) in Figure~\ref{fig:sym3}. These results are consistent with the premise upon which the data is generated.

\begin{figure}
\centering
\subfigure[EMDB 1654]{\label{fig:sym4}\includegraphics[width=0.48\textwidth]{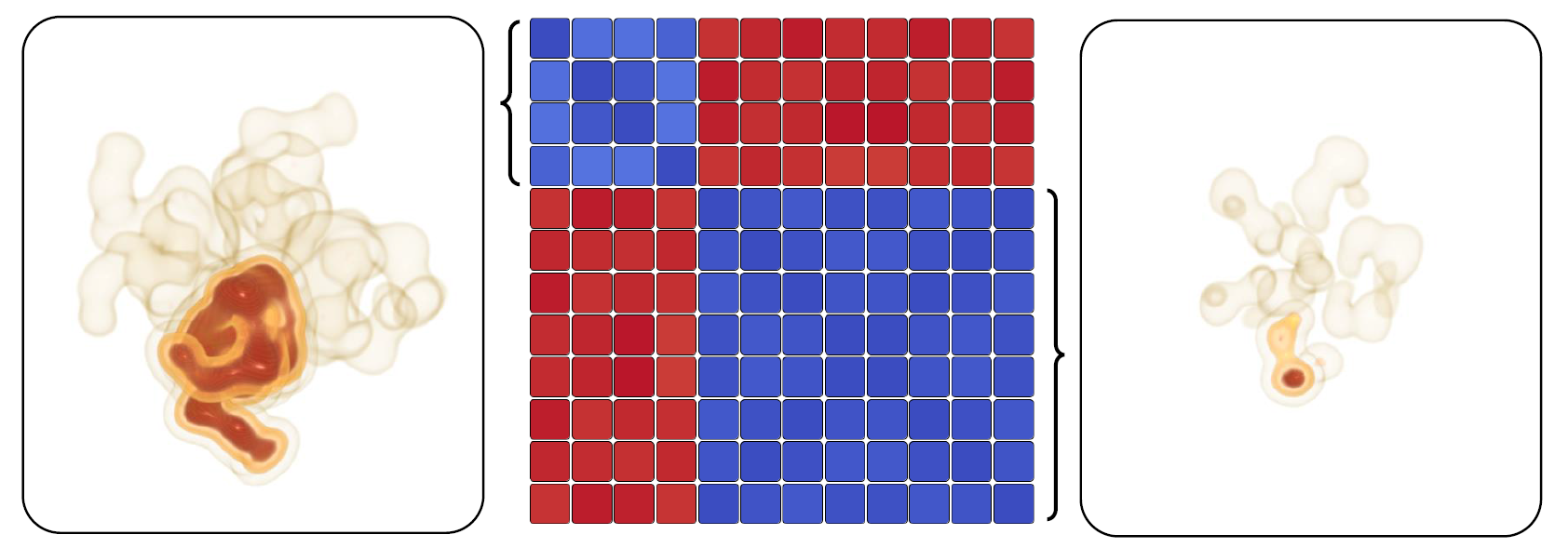}}
\subfigure[EMDB 1897]{\label{fig:sym5}\includegraphics[width=0.48\textwidth]{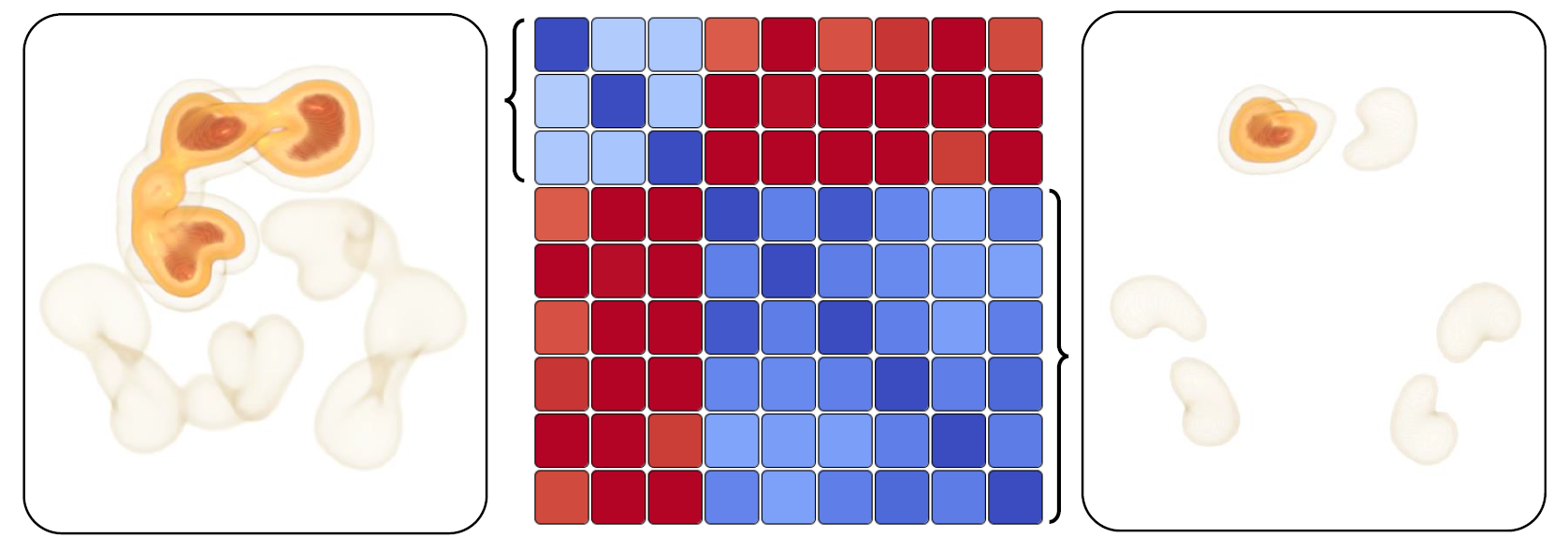}}
\caption{Detecting groups of symmetric regions in EMDB datasets. (centre)~DM showing tree edit distance between various pairs of subtrees of the merge tree. Low values are mapped to blue and high values to red. The DM indicates the presence of two distinct groups. All regions within a group are symmetric copies of each other. (left, right) Volume rendering where one region from each symmetric group is highlighted.
}
\label{fig:symm2}
\vspace{-0.15in}
\end{figure}

We present additional case studies that demonstrate the applicability of the tree edit distance measure to symmetry identification. EMDB\footnote{\url{https://www.ebi.ac.uk/pdbe/emdb/}} contains 3D electron microscopy density data of macromolecules, subcellular structures, and viruses. Some of these structures contain symmetric subunits. We study two structures, EMDB~1654, and~1897, see Figure~\ref{fig:symm2}. First, we compute the split tree for each structure. We then use a semi-automated method to extract sub-trees corresponding to significant features from the merge tree based on user specified persistence and minimum scalar value thresholds. Next, we compute the tree edit distance measure between sub-trees corresponding to these regions of interest. We observe two distinct groups from the DMs. The tree edit distance measure clearly identifies two groups with $4$ and $8$ regions each in EMDB~1654, see Figure~\ref{fig:sym4}. Similarly, it identifies two groups containing $3$ and $6$ symmetric regions each in EMDB~1897, see Figure~\ref{fig:sym5}.

\subsection{Shape matching}

\begin{figure}
\centering
\includegraphics[width=0.45\textwidth]{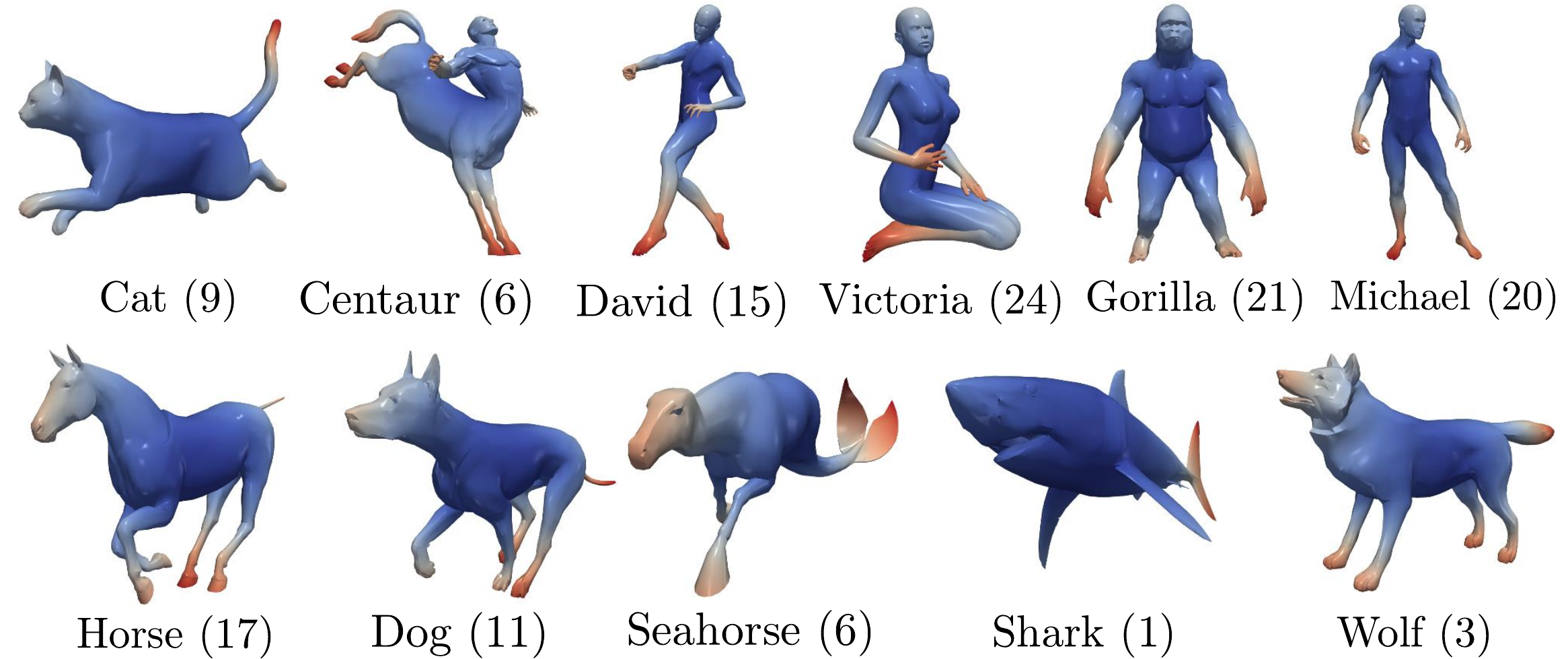}
\caption{Collection of shapes from the TOSCA non-rigid world dataset. The average geodesic distance field~\cite{hilaga2001} is computed on the surface. Each shape is available in multiple poses (number of poses mentioned within parenthesis), only one pose is shown here.}
\label{fig:shapes}
\vspace{-0.15in}
\end{figure}

\begin{figure}
\centering
\includegraphics[width=0.48\textwidth]{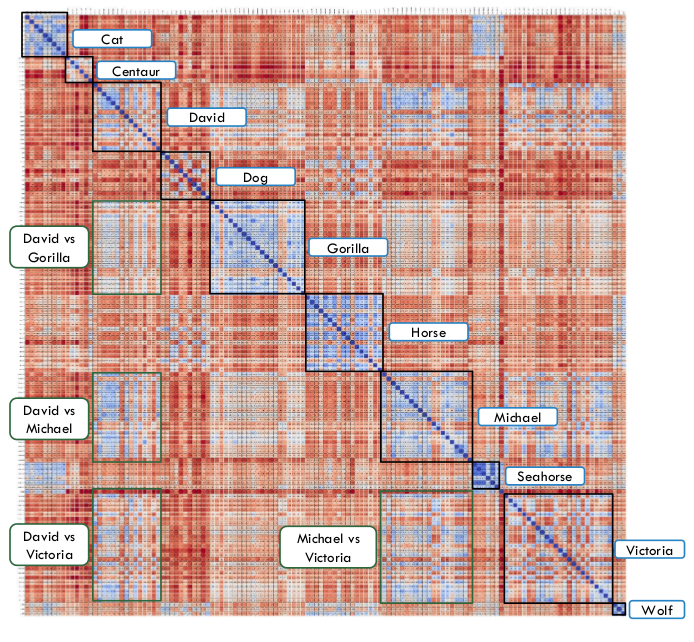}
\caption{Tree edit distance matrix for all pairs of shapes from the TOSCA non-rigid world dataset. Blocks of low values (blue) correspond to similar shapes but in different poses.}
\label{fig:mat}
\vspace{-0.15in}
\end{figure}

Shape matching involves comparing geometric shapes and finding similarity between them. A good distance measure helps quantify this notion of similarity more concretely. The TOSCA non-rigid world dataset\footnote{\url{http://tosca.cs.technion.ac.il/book/resources_data.html}} contains a set of different shapes, see Figure~\ref{fig:shapes}. The shapes are in different poses and the project aims to develop methods to identify similarity between shapes in a pose invariant manner. We compute the average geodesic distance field~\cite{hilaga2001} on the surface mesh. This field is well studied in the literature and is known to be a good shape descriptor. We apply a persistence simplification threshold of 1\% on the merge trees both to remove topological noise and to reduce the number of nodes. Next, we compute the tree edit distance measure between all pairs of shapes. It takes around $15$ seconds to generate the distance matrix with the same setup used for the periodicity experiment.
Figure~\ref{fig:mat} shows the distance matrix. Each collection of shape appears as a blue block irrespective of variations in pose.
We also observe higher values for a pair of shapes that are different. Note the blue blocks away from the diagonal. They correspond to Michael \textit{vs} Victoria, David \textit{vs} Victoria, David \textit{vs} Michael, and David \textit{vs} Gorilla. These pairs have similar shapes, which is more apparent in a few poses. Not all poses are shown in Figure~\ref{fig:shapes}.

\subsection{Data Summarization}
Exploring large scientific data, particularly time-varying data, and identifying patterns of interest is often time consuming even with good visualization tools. Well designed abstract representations provide good overviews of the data and direct the user to features of interest. Abstractions such as the merge trees present a summary of spatial features. Temporal summaries enable effective visualization of time-varying data. Central to the design of a temporal summary is a good distance measure that can distinguish between periods of significant activity and inactive time periods.

\begin{figure}
\centering
\includegraphics[width=0.45\textwidth]{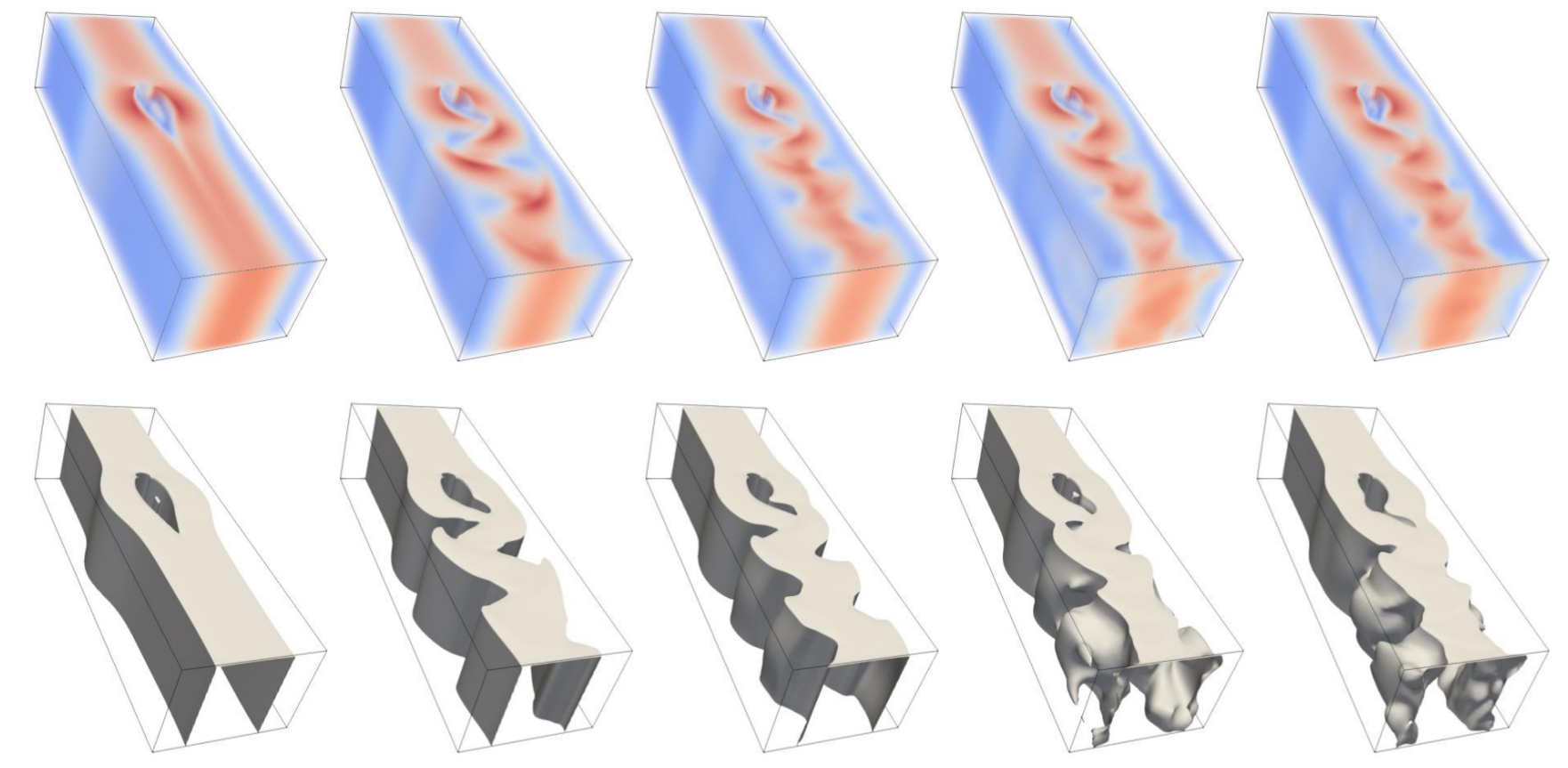}
\caption{The 3D B\'enard-von K\'arm\'an vortex street dataset. (top)~Volume rendering of the velocity magnitude field for time steps $15,35,58,91,98$ ordered left to right. (bottom)~Isosurfaces at isovalue $0.7$ extracted for the above time steps.}
\label{fig:3dvort}
\vspace{-0.1in}
\end{figure}

In this experiment, we consider the 3D B\'enard-von K\'arm\'an vortex street dataset. The velocity magnitude is available as a scalar field on a $192 \times 64 \times 48$ grid over $102$ time  steps~\cite{vonfunck08a}. Figure~\ref{fig:3dvort}  shows volume renderings and isosurfaces for a few time steps. Topological features of the velocity magnitude scalar field are represented using the split tree. Each split tree had approximately $180-200$ nodes. We compute the tree edit distance between all pairs of time steps. It takes around $4$ seconds to generate the distance matrix (DM) with the same setup used for the periodicity experiment.

\begin{figure}
\centering
\includegraphics[width=0.48\textwidth]{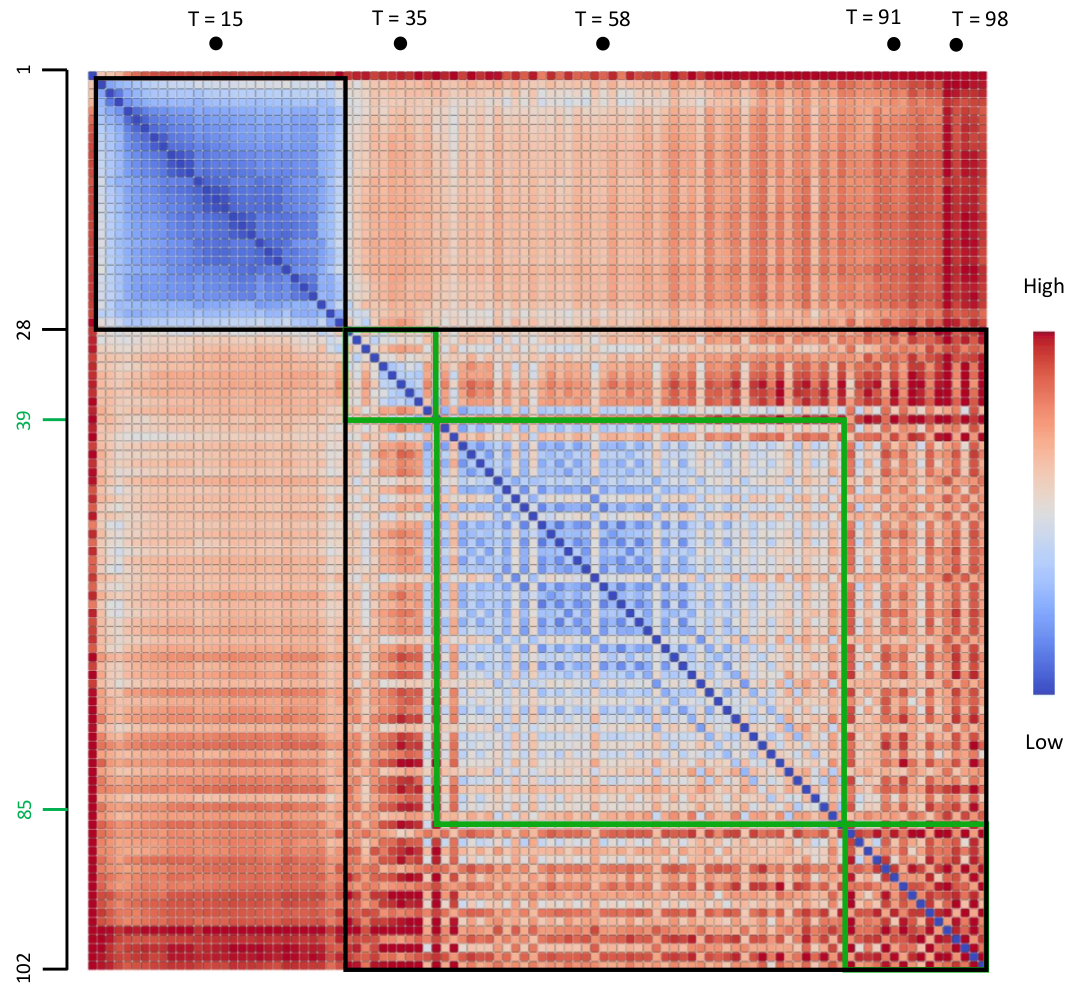}
\caption{Tree edit distance matrix for all time steps of the 3D B\'enard-von K\'arm\'an vortex street dataset. Columns corresponding to time steps $15,35,58,91,98$ are highlighted. Patterns that help in generating a temporal summary are highlighted using black and green boxes.}
\label{fig:dm}
\vspace{-0.2in}
\end{figure}

The DM shown in Figure~\ref{fig:dm} contains multiple patterns. A fluid dynamics expert helped study and interpret the results.  The distance between time steps $2-28$ are small because the flow does not contain any vortices and the features do not change. The top left blue block in the matrix corresponds to this time period. This is followed by the period when new vortex structures are formed (small block highlighted in green that corresponds to time steps $29-39$). Next, the vortices exhibit shedding, which is shown by the repeating patterns present in the larger green block in the matrix (time steps $40-85$). Finally, the vortices are significantly distorted, which is captured by the high values of distance in the bottom right block. Thus we can use the patterns that emerge in the distance matrix to distinguish between different types of behavior and summarize the scientific phenomena using these patterns. 
\section{Conclusions}
We described a distance measure between two scalar fields that compares their merge trees. The distance measure is defined as the minimum cost of a set of restricted edit operations that transforms one tree into another. The edit operations and the associated costs are both intuitive and mathematically sound. The measure satisfies metric properties, can be efficiently computed, and is useful in practice. We study the properties of the measure and demonstrate its application to data analysis and visualization using various computational experiments. In future work, we plan to develop a theoretical analysis of the stability properties of the measure. Developing a comparative visualization framework based on the tree edit distance measure is also an interesting problem with potential applications to time-varying data and multifield data visualization. 


%

%
%
%
\ifCLASSOPTIONcompsoc
\section*{Acknowledgments}
\else
\section*{Acknowledgment}
\fi
This work is supported by the Department of Science and Technology, India (DST/SJF/ETA-02/2015-16), and Joint Advanced Technology Programme, Indian Institute of Science (JATP/RG/PROJ/2015/16), and the Robert Bosch Centre for Cyber Physical Systems, Indian Institute of Science. We thank Shrisha Rao for discussions on the data summarization experiment.



\bibliographystyle{IEEEtran}
\bibliography{paper}

\vskip -3\baselineskip plus -1fil

\begin{IEEEbiography}[{\includegraphics[width=1in,height=1.25in,clip,keepaspectratio]{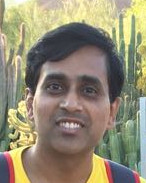}}]{Raghavendra Sridharamurthy} is a PhD candidate in computer science at Indian Institute of Science, Bangalore. He received BE degree in information technology from National Institute of Technology Karnataka, Surathkal and M.Sc degree in computer science from Indian Institute of Science. His research interests include scientific visualization, computational topology and its applications.
\end{IEEEbiography}

\vskip -3\baselineskip plus -1fil

\begin{IEEEbiography}[{\includegraphics[width=1in,height=1.25in,clip,keepaspectratio]{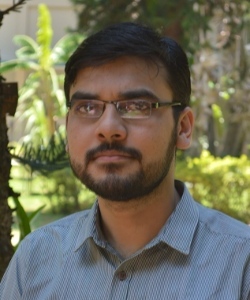}}]{Talha Bin Masood} is a PhD candidate in computer science at Indian Institute of Science, Bangalore. He received B.Tech degree from Aligarh Muslim University and ME degree in computer science from Indian Institute of Science. His research interests include scientific visualization, computational geometry, computational topology and its applications to various scientific domains.
\end{IEEEbiography}

\vskip -3\baselineskip plus -1fil

\begin{IEEEbiography}[{\includegraphics[width=1in,height=1.25in,clip,keepaspectratio]{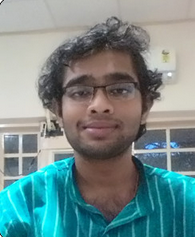}}]{Adhitya Kamakshidasan} is a Junior Research Fellow at Visualization and Graphics Lab, Indian Institute of Science. He holds a B.Tech degree in computer science from Visvesvaraya National Institute of Technology, Nagpur. His research interests include fluid simulation, information visualization and cartography.
\end{IEEEbiography}

\vskip -3\baselineskip plus -1fil

\begin{IEEEbiography}[{\includegraphics[width=1in,height=1.25in,clip,keepaspectratio]{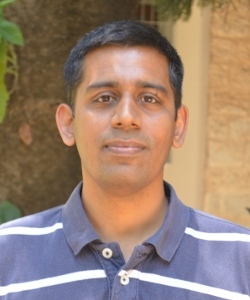}}]{Vijay Natarajan} is an associate professor in the Department of Computer Science and Automation at the Indian Institute of Science, Bangalore. He received the Ph.D. degree in computer science from Duke University in 2004. His research interests include scientific visualization, computational topology, and geometry processing.
\end{IEEEbiography}

\end{document}